\documentclass[%
 reprint,
 twocolumn,
superscriptaddress,
 amsmath,amssymb,
 aps,
 pra,
]{revtex4-2}
\usepackage[dvipsnames]{xcolor}

\newcommand{\gud}[0]{g_{\uparrow\downarrow}}
\usepackage{graphicx}% Include figure files
\usepackage{dcolumn}% Align table columns on decimal point
\usepackage{bm}% bold math
\usepackage{hyperref}% add hypertext capabilities
\usepackage{comment}
\usepackage{lipsum}
\usepackage{soul}

\begin{document}

\preprint{APS/123-QED}

\title{Quantum Monte Carlo-based density functional for one-dimensional Bose-Bose mixtures}
\author{Jakub Kopyci\'{n}ski}
\email{jkopycinski@cft.edu.pl}
\affiliation{Center for Theoretical Physics, Polish Academy of Sciences, Al. Lotnik\'{o}w 32/46, 02-668 Warsaw, Poland}
\affiliation{Joint Quantum Centre (JQC) Durham–Newcastle,\\
School of Mathematics, Statistics and Physics, Newcastle University,\\
Newcastle upon Tyne, NE1 7RU, United Kingdom}
\author{Luca Parisi}
\affiliation{Joint Quantum Centre (JQC) Durham–Newcastle,\\
School of Mathematics, Statistics and Physics, Newcastle University,\\
Newcastle upon Tyne, NE1 7RU, United Kingdom}
\author{Nick G. Parker}
\affiliation{Joint Quantum Centre (JQC) Durham–Newcastle,\\
School of Mathematics, Statistics and Physics, Newcastle University,\\
Newcastle upon Tyne, NE1 7RU, United Kingdom}
\author{Krzysztof Paw{\l}owski}
\affiliation{Center for Theoretical Physics, Polish Academy of Sciences, Al. Lotnik\'{o}w 32/46, 02-668 Warsaw, Poland}

\date{\today}

\begin{abstract}

We propose and benchmark a 
Gross-Pitaevskii-like equation for two-component Bose mixtures with competing interactions in 1D. Our approach follows the density functional theory with the energy functional 
based on the exact quantum Monte Carlo (QMC) simulations. Our model 
covers, but goes beyond, the popular approach with the Lee-Huang-Yang corrections.
We first benchmark our approach against available QMC data in all interaction regimes and then study dynamical properties, inaccessible by {\it ab initio} many-body simulations.
Our analysis includes a study of monopole modes  
and reveals the presence of anomalous dark solitons.

\end{abstract}

\maketitle

\section{Introduction}

Recent studies of ultracold gases with competing interactions have led to a major change in the field. They undermined the validity of the mean-field approximation 
when attractive and repulsive interactions in the system almost cancel each other.

Although at first one may think it is due to three-body interactions like it was described in one of the earliest works on droplets~\cite{Bulgac_2002},
it is in fact necessary to include the effect of quantum fluctuations~\cite{Petrov_2015}. To account for it, one 
can include Lee-Huang-Yang (LHY) corrections ~\cite{Lee_1957a, Lee_1957b} to the mean-field equation using a local density approximation. One then derives 
a generalized Gross-Pitaevskii (GGP) equation.
It has been widely used to theoretically investigate the ground-state properties and excitations of Bose-Bose mixtures with particular attention given to the 
compressional mode (known also as the monopole or breathing mode)~\cite{Astrakharchik_2018, Petrov_2016, Tylutki_2020, Flynn_2022}.
The GGP theory predicts the existence of self-bound objects -- ultradilute quantum droplets 
made of ultracold atoms~\cite{Petrov_2015}. The emergence of a liquid phase is marked by the presence of a local minimum in the energy density functional.

Soon after having been proposed theoretically, quantum droplets were experimentally observed ~\cite{Cabrera_2018, Cheiney_2018, Semeghini_2018, Ferioli_2019, DErrico_2019}. Some theoretical predictions indicate even the possibility of finding 
quantum droplets~\cite{Bisset_2021} in recently obtained heteronuclear dipolar condensates~\cite{Trautmann_2018,Durastante_2020}.
Despite its remarkable usefulness, there are still factors not included in the GGP theory. For instance, \textit{ab initio} calculations show a liquid-gas transition in two-component mixtures~\cite{Parisi_2019}, whereas the GGP does not. Moreover, the same work demonstrates a quantitative disagreement of the homogeneous state energy. Quite unexpectedly, the monopole mode frequencies happen to match the QMC calculations, though~\cite{Parisi_2020}. The nature of the liquid-gas transition still remains an open question. It is an especially interesting in the light of the Mermin-Wagner theorem~\cite{Mermin_1966, Hohenberg_1967}. Unfortunately, such a question cannot be answered in a purely numerical model we are about to present.

Several attempts have been made to overcome the existing imperfections of the GGP equation. One of the ideas, that follows the density functional theory, was to build an equation which would quantitatively reproduce the spatially uniform state energy from a chosen \textit{ab initio} method for any interaction strength.
In this regard the 1D Bose contact gas is a special system as its ground-state energy has been already derived in the analytical \textit{ab initio} calculations by E. Lieb and E. Liniger~\cite{Lieb_1963, Lieb_1963b}.

Using this exact energy functional one gets the single-particle equation here referred to as the Lieb-Liniger Gross-Pitaevskii (LLGP) equation that was used in Refs.~\cite{Dunjko_2001, Ohberg_2002, Kim_2003, Damski_2004, Damski_2006, Peotta_2014,Choi_2015, Kopycinski_2022b}. 
The equation proved to  correctly describe the ground state and low-lying excitations in all regimes -- from the weakly-interacting one (which, contrary to the 3D case happens at high gas densities) up to the Tonks-Girardeau regime (at low gas densities). The LLGP equation was recently used to study Bose gas with repulsive short-range and attractive dipolar interactions \cite{Oldziejewski_2020, DePalo_2021, Kopycinski_2022a, DePalo_2022, Lebek_2022} to show the existence and properties of the dipolar quantum droplets.
Concerning the droplets in quantum mixtures, a  
similar approach was employed to construct a quantum Monte Carlo (QMC)-based energy density functional for bosonic mixtures in 3D~\cite{Cikojevic_2020}, but so far the 1D Bose mixture was not investigated in such framework. For the latter system it was shown \cite{Parisi_2019}  that GGP fails to reproduce the phase diagram in certain regimes, in particular at low densities when atoms bind together into interacting dimers.

In this article, we aim to formulate and benchmark 
the QMC-based single-orbital density functional theory that is applicable to two-component Bose mixtures with repulsive intra- and attractive intercomponent interactions. We later refer to it as Lieb-Liniger Gross-Pitaevskii for mixtures (mLLGP).
Our theoretical approach using a single orbital $\psi$ and a QMC-based energy density functional $\mathcal{E}$ results in an equation of the following form:
\begin{equation}
\begin{split}
    i\hbar\partial_t\psi(x,t)=-\frac{\hbar^2}{2m}\partial^2_x\psi(x,t)+\frac{\delta \mathcal{E}}{\delta n}\psi(x,t),
    \label{eq:gen_DFT}
\end{split}
\end{equation}
where $n$ is the particle density. We want it to be applicable to two-component Bose mixtures with repulsive intra- and attractive intercomponent interactions.
To do this, we analyse the phase diagram of the system and numerically study the static properties and monopole mode of quantum droplets. We use data from Ref.~\cite{Parisi_2019} to construct the energy density functional and Ref.~\cite{Parisi_2020} to benchmark our approach.

In lower-dimensional systems, we can name two substantial beyond-LHY approaches. One of them is a pairing theory for bosons~\cite{Hu_2020}. The other one is based on the inclusion of higher-order corrections to the GGP equation~\cite{Ota_2020}. Both of them generally give only a qualitative agreement with QMC calculations. 

Our approach shares similarities with a density functional theory~\cite{Fiolhais_2003} for Fermi systems at unitarity~\cite{Bulgac_2012a}. The resulting density functional has been employed multiple times to look into strongly interacting fermions~\cite{Bulgac_2009, Wlazlowski_2018, Magierski_2019, Tylutki_2021, Magierski_2022}, revealing a remarkable consistency with the experiments~\cite{Zwierlein_2006, Ku_2016}.

A great advantage of having a Gross-Pitaevskii-like equation, in comparison to the QMC methods, is the possibility of studying nonlinear and time-dependent effects like the existence of dark solitons. This subject is particularly interesting as we may expect fundamentally different results than the solitons we know from single-species systems~\cite{Jackson_1998} or dark-dark solitons occurring in miscible bosonic mixtures~\cite{Morera_2018, Keverekidis_2016}. 

Very recently there have been reports on wide soliton-like objects, both in mixtures~\cite{Edmonds_2022} and in dipolar Bose gases~\cite{Kopycinski_2022b}. As such, last but not least, we show density and phase profiles of solitary waves
evaluated with our theory.

\section{Framework}

\subsection{System}

We consider a one-dimensional Bose gas consisting of two components $\sigma=\{\uparrow,\downarrow\}$ in a box of size $L$. We assume that the components have equal atomic masses $m_\uparrow=m_\downarrow=m$. We also assume that the short-range interaction coupling constants are the same in the intracomponent case $g_{\uparrow\uparrow}=g_{\downarrow\downarrow}=g$, whereas the intercomponent interactions can be independently tuned with a coupling constant $\gud$. Atoms of the same species repel each other while the intercomponent interactions are attractive. The binding energy of an atomic pair in vacuum $\varepsilon_b=-(m\gud^2/4\hbar^2)$ is a relevant energy scale in the system, while for the length scale we choose the intracomponent scattering length $a=2\hbar^2/mg$.  In experimental setups, such a system can be realised as a spin-balanced gas of a single bosonic isotope, where spins $\sigma$ correspond to two different hyperfine levels and the interaction strengths can be tuned with magnetic field via Feshbach interactions.

We assume that we are in the miscible regime. The single-component densities are locked according to the condition $n_\downarrow/n_\uparrow=\sqrt{g_{\uparrow\uparrow}/g_{\downarrow\downarrow}}$~\cite{Petrov_2015}, which

holds even in inhomogeneous cases.  In our system this implies that there are equal number of atoms in each component $N_\uparrow=N_\downarrow=N/2$ and that the single-component densities are half of the total density $n_\uparrow=n_\downarrow=n/2$.  If the system is homogeneous, the overall density is equal to $n=N/L$.

\subsection{Generalized Gross-Pitaevskii and quantum Monte Carlo approaches}

\begin{figure}[b]
\includegraphics{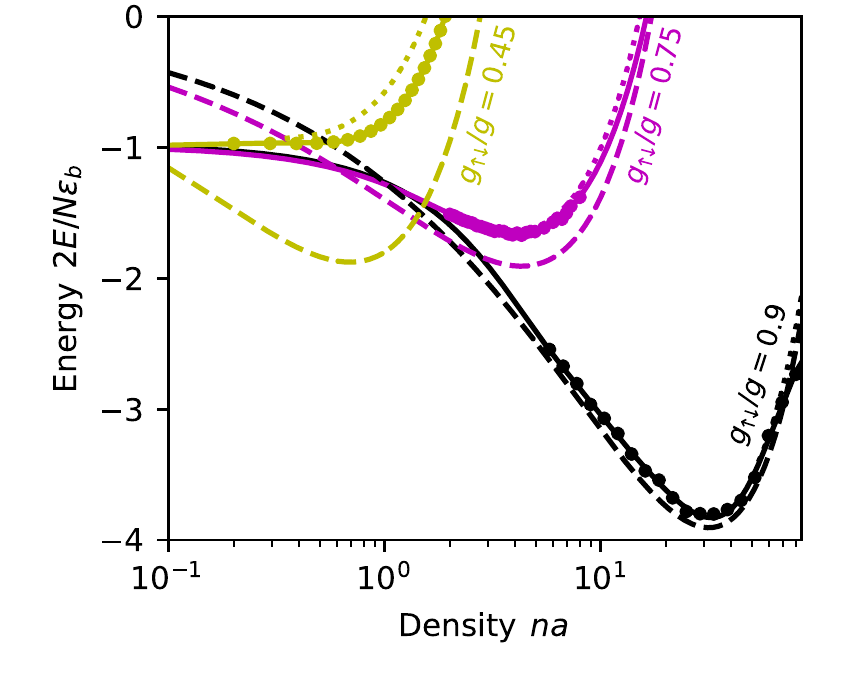}
\caption{Energy per particle as a function of density -- comparison of different models: QMC from Ref.~\cite{Parisi_2019} (markers), our density functional dubbed mLLGP (solid line), GGP (dashed line), pairing theory (dotted line) -  and interaction ratios $g_{\uparrow\downarrow}/g=0.45$ (yellow [light grey]), $0.75$ (magenta [grey]) and $0.9$ (black). The QMC error bars are smaller than the marker sizes.}
\label{fig:E_vs_dens}
\end{figure} 

In the weakly-interacting limit (corresponding to high densities $na\gg1$), one may expect the generalized GGP approach to be valid. The GGP energy density functional has the following part corresponding to interactions~\cite{Parisi_2019}:
\begin{equation}
\begin{split}
    \mathcal{E}_\mathrm{GGP}[n; g, \gud]=\frac{(g-\gud)n^2}{4}\\-\frac{mn^{3/2}}{3\sqrt{2}\pi\hbar}\left[(g-\gud)^{3/2}+(g+\gud)^{3/2}\right].
    \end{split}
    \label{eq:edf_ggp}
\end{equation}
The first term in Eq.~\eqref{eq:edf_ggp} corresponds to the mean-field contribution to the interaction energy and the other -- to the correction for quantum fluctuations, widely known as the LHY term.
If we compare, however, the results from GGP equation and \textit{ab initio} calculations from diffusion Monte Carlo in a wide range of densities and interaction ratios, we observe
discrepancies at low ratios. It is due to one of the peculiarities of one-dimensional systems -- the lower the density, the higher the interaction. Thus, the GGP model is correct in the high-density limit but cannot be trusted in the opposite case.

First of all, the GGP predicts the existence of stable quantum droplets for any ratio $\gud/g<1$. In other words, there is always a local minimum present in the energy density functional $\mathcal{E}_\mathrm{GGP}[n; g, \gud]$ as long as $\gud/g<1$. QMC predicts a certain critical value of the interaction ratio, below which the minimum disappears and we have a liquid-gas transition at $(\gud/g)_\mathrm{cr}=0.47(2)$~\cite{Parisi_2019}.

Although there are  
other methods, like a 
general extension to the LHY theory proposed in Ref.~\cite{Ota_2020}  
or a pairing theory for bosons introduced in~\cite{Hu_2020}, which are able to predict such a transition, they do not enable us to quantitatively compute the homogeneous gas energy with their use. Neither does the GGP, which results in an inaccurate estimate of a quantum droplet size and bulk density.

Lastly, the GGP is not applicable to the strongly-interacting regime. When $na\ll1$, the gas energy
quickly approaches half of the binding energy of a dimer,
i.e.,
$-\varepsilon_b/2$, indicating that the system could be understood as a weakly-interacting gas of dimers~\cite{Hu_2020}. The energy per dimer approaches~$-\varepsilon_b$ in the limit of vanishing density, while according to the GGP theory it tends to zero. 

\subsection{Lieb-Liniger Gross-Pitaevskii equation for two-component 1D bosonic mixtures (mLLGP equation)}

We aim to construct a novel 
energy density functional to study bosonic mixtures in 1D, which gives 
    (i) a quantitative agreement with QMC in terms of a homogeneous gas energy $E(n; g, \gud)$ in a wide range of interaction ratios~\footnote{Just as the original Lieb-Liniger Gross-Pitaevskii equation gives a quantitative agreement with the Lieb-Liniger model -- the homogeneous system energy, chemical potential and speed of sound are the same from construction.},
    (ii) a proper limit of a uniform gas energy, i.e. $\lim_{na\to0}E(n; g, \gud)=-N\varepsilon_b/2$,
    and (iii) a correct value for the critical interaction ratio $(\gud/g)_\mathrm{cr}$, at which a liquid-gas transition occurs.
It is 
more accurate than both the GGP and pairing theory, but, unlike QMC, enables us to study nonlinear and time-dependent effects, e.g. the properties of dark solitons.

To do that, we fit 
QMC data
from Ref.~\cite{Parisi_2019} to get a spline 
representation of the energy functional $\mathcal{E}_\mathrm{mLLGP}[n;g,\gud]$ and construct a single-orbital density functional theory for bosonic mixtures. To do this, we extrapolate the data in the low- and high-density regimes with two separate functions. This is necessary because the QMC data is covering only a part of densities, omitting the low- and high-density regions.
Afterwards, we interpolate the data with a spline in densities and linearly in interaction ratios. In this way, we obtain $\mathcal{E}_\mathrm{mLLGP}[n;g,\gud]$ in a form which is convenient for numerical evaluation. This whole procedure is described in detail in Appendix~\ref{app:splines}.

We decided to use a numerical representation of $\mathcal{E}_\mathrm{mLLGP}[n;g,\gud]$ after having checked a few simpler representations, including polynomials, but these representations did not fulfil the conditions (i)-(iii) we have listed earlier.

\begin{figure}[t]
\includegraphics{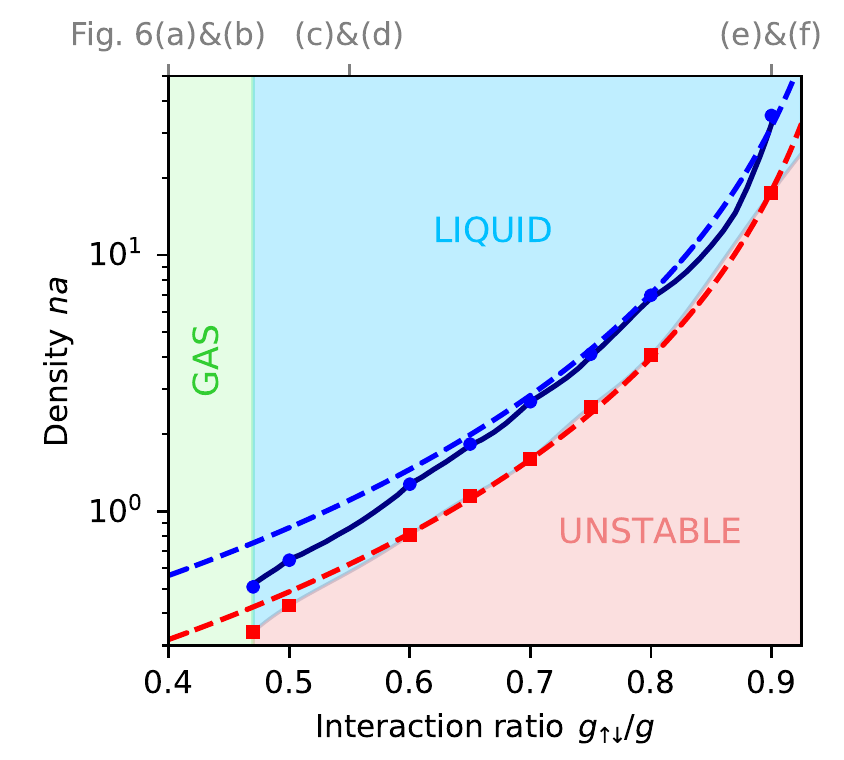}
\caption{Phase diagram of a homogeneous two-component mixture: the unstable region is demarcated by spinodal densities, predicted from GGP (red [light grey] dashed line) and QMC (square markers). Equilibrium density given by the mLLGP (navy [dark grey] solid line), GGP (blue [grey] dashed line) and QMC from Ref.~\cite{Parisi_2019} (round markers). The ticks on top correspond to the interaction ratios used in Fig.~\ref{fig:sol_params}.
}
\label{fig:phase_diag}
\end{figure}

Figure~\ref{fig:E_vs_dens} shows us the energy per particle of a homogeneous Bose-Bose mixture. For interaction ratios $\gud/g\simeq1$ all three theories (GGP, pairing theory and mLLGP) are consistent with QMC calculations. In the case of the GGP and pairing theory, the smaller the ratio becomes, the higher the discrepancy is. For ratio $\gud/g=0.45$, the energy per particle from the GGP model still possesses a pronounced minimum, whereas QMC, mLLGPE and the pairing theory predict a lack thereof. The latter deviates from the QMC data and matches it only qualitatively in this region. One can see the energy functional $\mathcal{E}_\mathrm{mLLGP}$ is constructed to fulfil all the conditions from the list above. 

The analysis of the energy functional in a  state can provide us with important thermodynamic quantities. For instance, $\mu_\mathrm{mLLGP}[n_0;g,\gud]=\delta\mathcal{E}_\mathrm{mLLGP}[n;g,\gud]/\delta n|_{n=n_0}$ is the chemical potential evaluated at density $n_0$, and the speed of sound $c$ is given by the following relation $c=\sqrt{\frac{n}{m}\frac{d\mu}{dn}}$. 
The position of the energy per particle minimum plays a vital role in the context of quantum droplet studies: the equilibrium density $n_\mathrm{eq}$ where $d(E/N)/dn=0$ is the value of the density in the droplet bulk, assuming the droplet is sufficiently large, i.e. $N\gg1$ and possesses a flat-top profile. In this limit, we may approximate the properties of the droplet bulk to be the same as those of a homogeneous system with density $n_\mathrm{eq}$.

\begin{figure}[b]
\includegraphics[width=\linewidth]{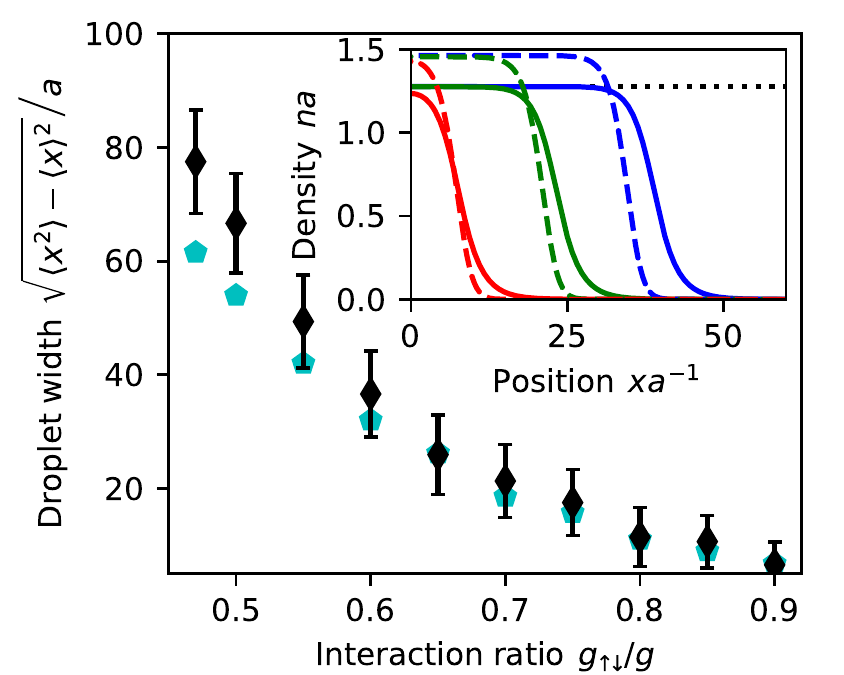}
\caption{Droplet width $\sqrt{\langle x^2\rangle-\langle x\rangle^2}$ as a function of the interaction ratio $g_{\uparrow\downarrow}/g$. Pentagonal cyan markers correspond to the GGP prediction, diamond black ones -- to the mLLGP estimation. Number of particles forming the droplet $N=100$. \\Inset: density profiles of quantum droplet evaluated at a ratio $g_{\uparrow\downarrow}/g=0.6$ using the mLLGPE (solid) and GGP (dashed) for different number of particles $N=20$ (red [innermost]), $60$ (green [middle]) and $100$ (blue [outermost]). Black dotted line corresponds to the equilibrium density given by the QMC calculations from Ref.~\cite{Parisi_2019}.}
\label{fig:dens_prof}
\end{figure} 

With that knowledge we are able to explore the phase diagram and compare it to the one created with the QMC approach. We show it in Fig.~\ref{fig:phase_diag}. We are able to distinguish 3 phases: gaseous, liquid and unstable. The gaseous one corresponds to the region where 
the minimum in the energy density functional is located at the vanishing density. It happens when the interaction ratio $\gud/g<0.47$. Above that value, the minimum exists and we enter the liquid phase. Nevertheless, in the region $\gud/g>0.47$, there is a range of densities for which the speed of sound is complex. This
signals a phonon instability. 

The unstable and stable liquid phases are demarcated by spinodal densities $n_\mathrm{ins}$, where $d^2\mathcal{E}/dn^2=0$. At this border, the compressibility is infinite. The nature of the unstable liquid phase in a weakly interacting bosonic mixtures was discussed in Ref.~\cite{DeRosi_2021}.
In Fig.~\ref{fig:phase_diag} we also plot equilibrium densities $n_\mathrm{eq}$ (see solid navy line for mLLGP and a dashed blue one for GGP). The two comparisons QMC vs GGP and QMC vs mLLGP favour the latter approach. Wherever we have data from QMC simulations, the mLLGP predicts the same equilibrium density as \textit{ab initio} calculations~\footnote{It is a benchmark of a correctly prepared fit.}. On the other hand, the GGP extends both liquid and unstable regions far beyond the critical interaction ratio $(\gud/g)_\mathrm{cr}$.

For low interaction ratios $\gud/g\ll1$, the equilibrium densities are located in the low-density region. However, in this limit of densities, the gas cannot be treated anymore as weakly-interacting. The GGP approach, contrary to QMC, gives us a rough estimate of $n_\mathrm{eq}$ only.

Having established that the constructed energy functional reproduces the phase diagram according the QMC theory, we can now use this to construct an equation of the form of Eq.~\eqref{eq:gen_DFT} which allows for modelling time dependence and inhomogeneity of the effective single particle orbital.  We now write this equation as:

\begin{equation}
\begin{split}
    i\hbar\partial_t\psi(x,t)=-\frac{\hbar^2}{2m}\partial^2_x\psi(x,t)\\+\mu_\mathrm{mLLGP}\left[|\psi(x,t)|^2;g,\gud\right]\psi(x,t).
    \label{eq:mLLGPE}
\end{split}
\end{equation}

The square modulus of this orbital is interpreted as the particle density $n(x)$. Next, in Sec.~\ref{sec:static}, we will numerically solve the mLLGP equation~\eqref{eq:mLLGPE}, with the use of imaginary time propagation to find broken-symmetry states in Bose-Bose mixtures. Following this, in Sec.~\ref{sec:dynamic} we will additional solve the equation in real time to simulate the breathing modes of a perturbed droplet.  Our toolkit is provided under the link \url{https://gitlab.com/jakkop/mudge/-/releases/v07Mar2023}.

\section{Quantum droplets}
\subsection{Static properties\label{sec:static}}
The ground state (GS) of a two-component mixture in the liquid regime takes a form of a quantum droplet. Typical density profiles of one-dimensional droplets are shown in the inset of Fig.~\ref{fig:dens_prof}. The quantum droplets evaluated with the mLLGP (see Appendix~\ref{app:ITE_RTE} for numerical details) exhibit a flat-top bulk when the number of particles exceeds $20$. For $N=60$ and $100$, we can observe a prominent plateau with the same density as the equilibrium value $n_\mathrm{eq}$ given by QMC calculations. We juxtaposed these density profiles with analogous ones given by the GGP equation. As we can see, their bulk densities do not match the QMC prediction. The discrepancy for $\gud/g=0.6$ is equal to $14\%$, but grows up to $48\%$ at the critical ratio $(\gud/g)_\mathrm{cr}=0.47(2)$ (cf.\ Fig.~\ref{fig:phase_diag}).

As the number of particles in the droplet $N$, its bulk density $n_\mathrm{eq}$ and its width $\sqrt{\langle x^2\rangle-\langle x \rangle^2}$ are connected (in the first approximation $n_\mathrm{eq}\propto N/\sqrt{\langle x^2\rangle-\langle x \rangle^2}$), the difference between the estimations of the equilibrium density should be also visible when we plot the droplet width against the interaction ratio, but keeping a fixed number of atoms in the system. We show it in the main panel of Fig.~\ref{fig:dens_prof}.

As we can see, the droplet width is a decreasing function of $\gud/g$. When $\gud$ becomes larger, the interparticle attraction gets more pronounced and the droplet contracts. As expected, the GGP gives a qualitative agreement of the droplet width with the mLLGP. However, the lower the interaction ratio, the higher the discrepancy between the models.

\begin{figure}[h!]
\includegraphics[width=\linewidth]{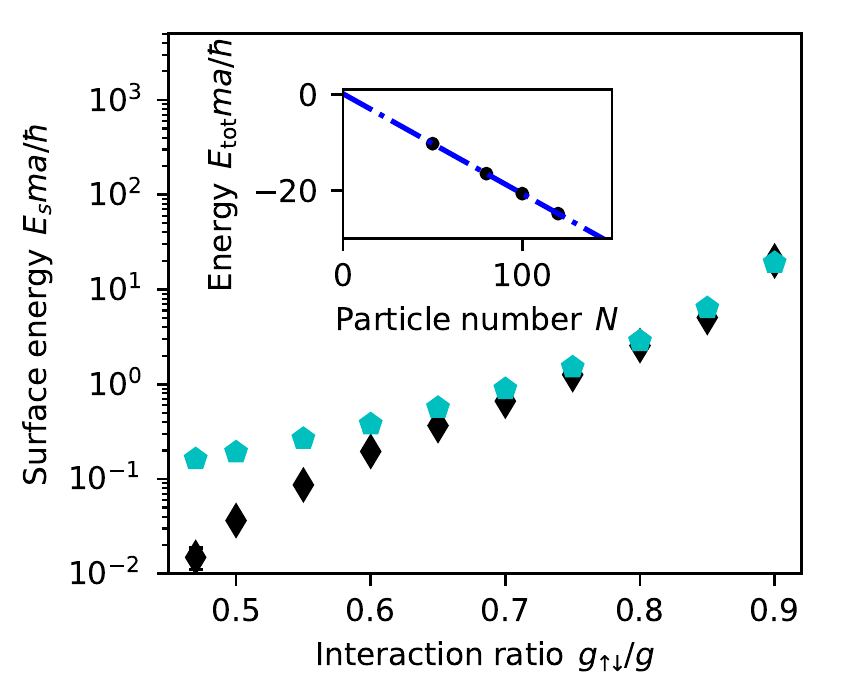}
\caption{Surface energies of quantum droplets for different interaction ratios $\gud/g$. Pentagonal cyan markers correspond to the GGP results, diamond black ones -- to the mLLGP prediction.\\
Inset: total energy of a quantum droplet obtained with Eq.~\ref{eq:mLLGPE} for high particle numbers $N\gg1$ and $\gud/g=0.6$. The dash-dotted line corresponds to a linear fit $E_\mathrm{tot}=\overline{e}N+E_S$.\\
The main contribution to the uncertainty of the surface energy is due to the linear fit and in most cases, the error bars are smaller than the marker size.}
\label{fig:surf}
\end{figure} 

In classical physics the total energy of the droplet can be divided into the volume and surface terms $E_\mathrm{tot}=E_V+E_S$. In a one-dimensional system, the surface term should be $N$-independent and the volume term (for $N\gg1$) should be proportional to the number of particles in the droplet as we show it in the inset of Fig.~\ref{fig:surf}. We are particularly interested in the value of the surface term. If a droplet gets split, the energy in the system increases by $E_S$. Low values of the surface energy may be considered an issue in the experiment. Namely, thermal excitations might cause a fission of the droplet.

Figure~\ref{fig:surf} depicts the surface energy of the droplet as a function of the interaction ratio $\gud/g$. In the case of mLLGP, the diminishing surface energy when approaching $(\gud/g)_\mathrm{cr}$ is a signature of the liquid-gas transition proximity. The surface tension slowly decreases until it vanishes below the critical interaction ratio. The GGP does not predict such a transition, so the surface tension does not go to zero according to this theory.

One may ask here on the contribution from gradient corrections to the energy functional and their influence on the results. As the analysis conducted in Ref.~\cite{Parisi_2020} shows a quantitative agreement of the surface energy of the droplet between the QMC and GGP in the weakly interacting regime, we do not include them into $\mathcal{E}_{\rm mLLGP}$.

\subsection{Monopole mode excitation\label{sec:dynamic}}

\begin{figure}[t!]
\includegraphics[width=\linewidth]{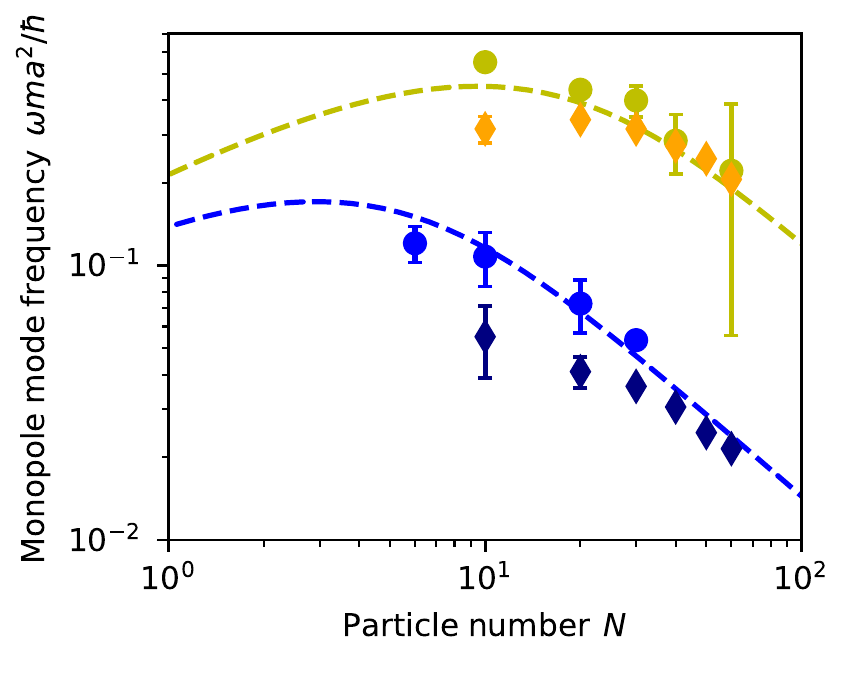}
\caption{Monopole mode frequency as a function of the number of particles in the droplet. Round markers correspond to the linear response theory prediction based on QMC data from Ref.~\cite{Parisi_2020}, diamonds -- to the mLLGP, and the dashed lines -- to the GGP predictions. Frequencies evaluated at a ratios $g_{\uparrow\downarrow}/g=0.6$ (blue [dark grey]) and $0.8$ (yellow [light grey]).}
\label{fig:br_mode}
\end{figure}

We 
now look into how the ground state
reacts to a small perturbation.  We choose to study the monopole mode. We evolve in real time a quantum droplet perturbed by a factor $\exp(-i\epsilon x^2/a^2)$, where $\epsilon$ is a small constant. It corresponds to a situation when the initial velocity field in a droplet has the form $v(x)=-2\hbar\epsilon x/ma^2$ (further details are provided in Appendix~\ref{app:ITE_RTE}). At the beginning, the droplet is squeezed and at some point it expands again. This process is periodic and has its characteristic frequency which we measure by looking at the standard deviation of the droplet width $\sqrt{\langle x^2\rangle-\langle x\rangle^2}$ in time.

\begin{figure*}[t]
\includegraphics[width=\linewidth]{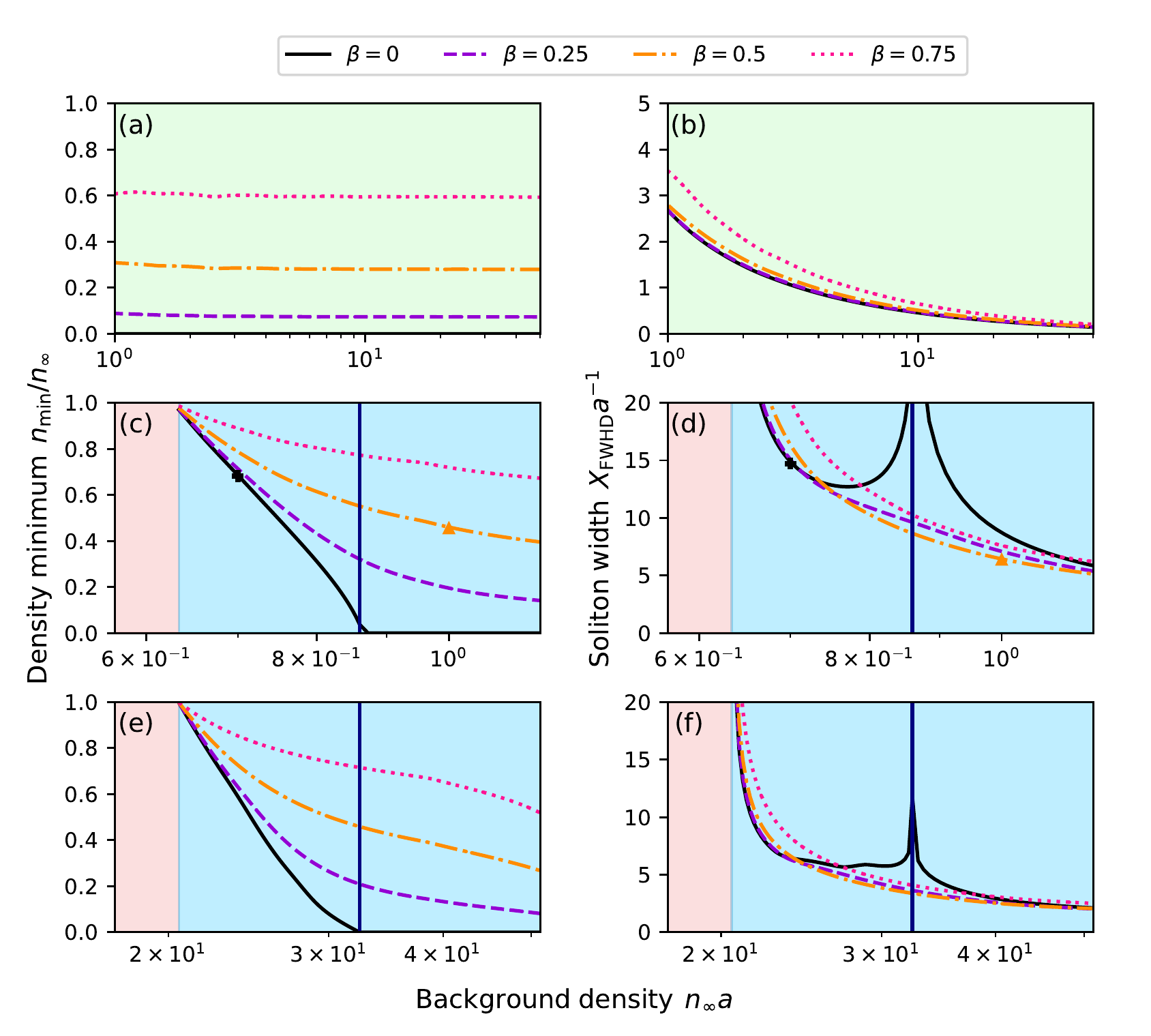}
\caption{Minimum densities $n_\mathrm{min}$ (a, c, e) and full widths at half depth $X_\mathrm{FWHD}$ (b, d, f) of dark solitons for different relative velocities of the soliton $\beta$ and interaction ratios $\gud/g=0.4$ (a, b), $0.55$ (c, d), and $0.9$ (e, f). Green shading corresponds to gaseous phase, blue -- to the liquid one and red -- to the unstable regime (cf.\ Fig.~\ref{fig:phase_diag}). The vertical blue line marks the equilibrium density value. The black plus-shaped markers in panels (c) and (d) correspond to the soliton shown in Fig.~\ref{fig:sol_prof}(a, c) and the orange triangles -- to the soliton from Fig.~\ref{fig:sol_prof}(b, d)}
\label{fig:sol_params}
\end{figure*}

We show the results of this numerical analysis in Fig.~\ref{fig:br_mode} altogether with the monopole mode frequencies evaluated with the GGP~\cite{Astrakharchik_2018} and linear response theory predictions based on QMC data~\cite{Parisi_2020}, i.e. the data which were not used in to fit $\mathcal{E}_\mathrm{mLLGP}$. All three approaches give consistent results in the large particle number limit. The monopole mode frequency scales like $\omega\propto N^{-1}$ there~\cite{Astrakharchik_2018}. Surprisingly, the QMC data also agree with the GGP-based results, even though the GGP equation 
is not expected to be accurate for small $N$, %\st{mostly} 
due to the 
breakdown of the local density approximation (LDA), which 
requires fulfilling the condition  $N\gg1$.

The mLLGP simulations agree in most cases within the range of 2 uncertainties. The dominating source of uncertainty is the form of $\mathcal{E}_\mathrm{mLLGP}$ in the low-density regions $n\ll n_\mathrm{eq}$. As we lack Monte Carlo data there, we cannot control the quality of the fit below the equilibrium density. It is clearly visible when the number of particles in the droplet is low. The bulk density is lower than $n_\mathrm{eq}$ there (cf.\ the inset of Fig.~\ref{fig:dens_prof}), especially after a slight expansion happening due to the perturbation we apply.

Thus, an accurate 
measurement of monopole mode frequencies 
seems to be the best choice to experimentally verify the validity of mLLGP-based study. It might be a daunting task, though. The difference is most striking in the small-droplet limit, which might be difficult to achieve in an experimental setup.

\section{Dark solitons}

We supplement our study of Bose-Bose mixtures with a numerical analysis of dark solitons. They are an example of nonlinear effects which are beyond the range of QMC.

\begin{figure*}[t]
\includegraphics[width=\linewidth]{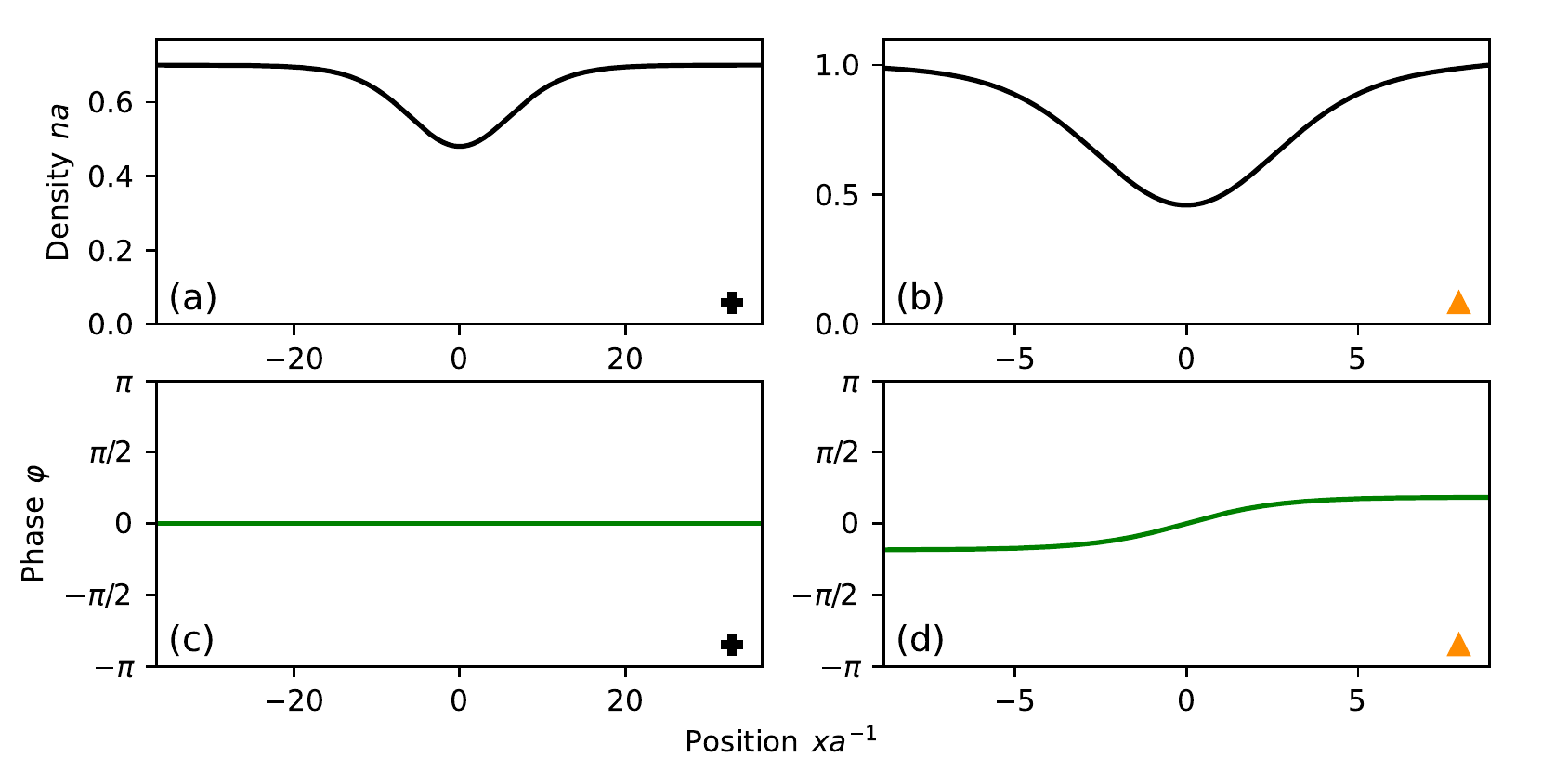}
\caption{Motionless anomalous soliton density (a) and phase (c) profiles. Standard grey soliton density (b) and phase (d). The grey soliton is moving with relative velocity $\beta=0.5$. Both solitons were evaluated at a ratio $g_{\uparrow\downarrow}/g=0.55$ using the mLLGPE.Black plus-shaped and orange triangle markers show a correspondence between this figure and Fig.~\ref{fig:sol_params}(c,d)}
\label{fig:sol_prof}
\end{figure*}

We look for solitonic solutions of the mLLGP equation~\eqref{eq:mLLGPE} in the thermodynamic limit. By dark soliton we understand a density depletion travelling at a constant velocity $v_s$ without changing its shape. We may classify these solitons as grey solitons if $v_s>0$ and they have a non-zero density minimum, and as black solitons if they are motionless and their density minimum is equal to zero~\cite{Jackson_1998}.

We assume that the density and phase of the orbital $\varphi=\arg \psi$ far from the soliton are constant and equal to $n_\infty$ and $\varphi_\infty$. Our numerical methods (see Appendix~\ref{app:solitons} for details) enable us to find both motionless and moving dark solitons. We use a velocity relative to the speed of sound $\beta=v_s/c$ to characterize the soliton.

If we take a look at the dark solitonic solutions in the weakly-interacting single-component Bose gas, we encounter both moving and motionless solutions.

Figure~\ref{fig:sol_params} presents the solitonic density minima $\min n(x)$ and full widths at half depths $X_\mathrm{FWHD}$ as functions of the density $n_\infty$ for three values of the interaction ratio $\gud/g$. The motionless solitons in the gaseous phase $n_\mathrm{eq}$ can be classified as standard ones 
-- their density reaches zero (cf.\ black solid line in panel (a)). Moreover, the density minima of grey solitons increase with their velocity. Panel (b) shows us the soliton width, which diverges as $n_\infty\to0$.

The situation changes when we cross the critical interaction ratio and enter the liquid phase.  In the high-density limit the soliton minimum density is zero, but below $n_\mathrm{eq}$ we enter a region where the 
minimum density starts to increase (see panels (c) and (e)). One may say the motionless solitons greyen. These solitons have been first described in Ref.~\cite{Kopycinski_2022b} and arise due to nonlinearities. The presence of a local minimum at a finite value of the density in $E/N$ plays a crucial role here. Due to their uncanny features, described at length later in this section, we call them anomalous. We have confirmed that these solutions maintain their form and phase profile during real time propagation in the presence of low-amplitude noise, confirming their stability.

The solitonic solution (as 
shown in panels (d) and (f)) widens in two places. Once when $n_\infty\to n_\mathrm{eq}$, both in the standard ($n_\infty\to n_\mathrm{eq}^+$) and anomalous ($n_\infty\to n_\mathrm{eq}^-$) regimes and another time, while approaching the instability region. The most interesting 
regime to realise experimentally is 
in the vicinity of $n_\mathrm{eq}$. The solitonic solutions there are both wide and deep, which may be easier to detect with \textit{in situ} imaging procedure.

Grey solitons also become shallower with decreasing density $n_\infty$. For $\beta>0$, it is a gradual change though (cf.\ panels (c) and (e)). Another difference is that the grey soliton width does not diverge when $n_\infty\to n_\mathrm{eq}$, it does so in the vicinity of the unstable regime only (cf.\ panels (d) and (f)).

In Fig.~\ref{fig:sol_prof}(a) and (c) we show the density and phase profiles of motionless solitons evaluated at a ratio $\gud/g=0.6$ and density fulfilling the inequality $n_\mathrm{ins}<n_\infty<n_\mathrm{eq}$. This soliton has a non-zero density minimum, normally characteristic to moving (grey) solitons. Moreover, there is no 
$\pi$-phase jump, as in a standard motionless (black) solitonic solution in the GPE~\cite{Jackson_1998}.

On the other hand, when $n_\infty>n_\mathrm{eq}$,
no anomalous solutions are found. In this regime, 
solitons 
are similar to standard 
dark solitons. We show density and phase profiles of a grey soliton moving with velocity $\beta=0.5$ in panels (b) and (d) of Fig.~\ref{fig:sol_prof}.

To gain some insight into the large width of the solitons when $n_\infty\approx n_\mathrm{eq}$, we shall consider again a homogeneous gas. We can define the pressure as $P=- dE/dL$. Above the value of $n_\mathrm{eq}$, the pressure is positive. But below the equilibrium density, the pressure becomes negative. Thus, if we break the symmetry in the system by rarefying the density in one point, the pressure will make the gas on the sides of the defect contract and form a structure with wide density depletion.

\section{Summary}

To conclude, we have presented a QMC-based single-orbital density functional theory for a two-component bosonic mixture in one dimension, which we call the mLLGP model. From construction, our approach provides a quantitative agreement in terms of the energy and chemical potential of a homogeneous state with the \textit{ab initio} QMC model from Ref.~\cite{Parisi_2020}.

We benchmark our equation by comparing the results with the original QMC data. This comparison shows the mLLGP can quantitatively predict the bulk density of a quantum droplet and the monopole mode frequency in the limit of a large number of particles in the droplet with a characteristic $\omega\propto N^{-1}$ dependency. It also predicts a correct phase diagram of Bose-Bose mixtures, including a transition from liquid to gas, not predicted by the mean-field model supplemented with the LHY correction. Since our approach relies on fitting an energy functional to QMC data, it is limited by the range of underpinning QMC data, which is only currently available in the literature for densities close to the equilibrium density and for specific interaction ratios.  Should QMC data become available over a larger parameter space of density and interaction ratio, the model could be refined with an improved energy functional.

Our work is limited to the specific case where the intraspecies interactions are equal, $g_{\downarrow\downarrow}=g_{\uparrow\uparrow}$, which leads to the density profile of each component being equal to each other, $n_\downarrow(x)=n_\uparrow(x)$.   In principle the approach could be extended to the more general case where $g_{\downarrow\downarrow}\neq g_{\uparrow\uparrow}$ and $n_\downarrow(x) \neq n_\uparrow(x)$, however this would require QMC data over a wider parameter space.  Given the computational intensity of QMC calculations, this is not tractable at the present time but may become possible in the future.

Lastly, we provide a brief study of solitonic solutions of the mLLGP equation, 
where we find ultrawide solitonic solutions. Moreover, 
anomalous motionless  
solitons were found 
as well. These solitons are characterized by the lack of a $\pi$-jump in the phase and a non-zero density minimum.

The presence of such wide solitons can be an advantage for experimenters who would like to perform an \textit{in situ} imaging of these objects. As far as we are concerned, the measurement of the monopole mode frequency for small droplets may be helpful to verify the validity of the mLLGP equation too. It would demand creating droplets consisting of very few particles, though, making such an experiment tougher to design and conduct. An avenue for further work would be to use the mLLGP model to study the dynamical properties of dark solitons in 1D Bose-Bose mixtures, particularly the anomalous solitons, including their collisions, stability and experimental generation.

\vspace{0.5cm}

\hypertarget{sec:data-avail}{\textit{Data availability ---}} All the numerical data necessary to reproduce figures, including QMC data and the results of simulations with the MUDGE toolkit (\url{https://gitlab.com/jakkop/mudge/-/releases/v07Mar2023}) are available in the Supplemental Material under the link [URL will be inserted by publisher].

\begin{acknowledgments}

The authors acknowledge discussions with Dr Thomas Billam and Mr Thomas Flynn (Newcastle University).

L.P. and N.P. acknowledge support from the UK Engineering and Physical Sciences Research Council (Grant
No. EP/T015241/1). J.K. and K.P. acknowledge support from the (Polish) National
Science Center Grant No. 2019/34/E/ST2/00289.

Center for Theoretical Physics of the Polish
Academy of Sciences is a member of the National Laboratory of Atomic, Molecular and Optical
Physics (KL FAMO).

L.P. prepared the energy density functional, J.K. conducted the numerical simulations. K.P. and N.P. conceptualized and supervised the research. J.K. wrote the manuscript with input of all authors.  

\end{acknowledgments}

\appendix
\section{{}\\Details of the numerical procedures~\label{app:splines}}

\textbf{Energy density functional}

In order to find the energy density functional $\mathcal{E}_\mathrm{mLLGP}[n;g,\gud]$, we use the QMC data from Ref.~\cite{Parisi_2019}, namely the energy per particle $E/N\equiv e(n;g,\gud)$ for the following interaction ratios $\gud/g=\{0.3, 0.4, 0.45, 0.5, 0.6, 0.65, 0.7, 0.75, 0.8, 0.9\}$. The data are extrapolated in the low density limit with a function $f_L(n)=-1+c_1n^{3/2}+c_2n^{5/2}+c_3n^3$ and $f_H(n)=c_4n^{1/2}+c_5n+c_6n^{3/2}$ [in units of $\varepsilon_b/2$], where $c_i$ for $i=\{1,2,\ldots, 6\}$ are constants to be fitted. Then, we perform a spline interpolation of the augmented QMC data and perform a linear interpolation between the ratios. The energy density functional is connected to the energy per particle function $e(n;g,\gud)$ via a simple relation: $\mathcal{E}_\mathrm{mLLGP}[n;g,\gud]=ne(n;g,\gud)$.

\textbf{Imaginary and real time evolution details\label{app:ITE_RTE}}

The mLLGP equation is a complex, nonlinear partial differential equation. The orbital $\psi(x)$ is discretized on a spatial mesh with $N_x$ fixed points and lattice spacing $DX=L /N_x$, where $L$ is the box size. We assume periodic boundary conditions, i.e. $\psi(-L/2)=\psi(L/2)$. The real-time evolution is done with the use of the split-step numerical method. The evolution with the kinetic term is done in the momentum domain, whereas the contact interaction term is calculated in the spatial domain. No external potential is used. The quantum droplet is obtained with the use of imaginary time evolution, where we use Wick rotation $t\to -i\tau$ to find the ground state.
The program written in C++ implementing the algorithm above is publicly available (see \hyperlink{sec:data-avail}{Data availability} for link).

The program
uses the W-DATA format dedicated to store data in
numerical experiments with ultracold Bose and Fermi
gases. The W-DATA project is a part of the W-SLDA
toolkit~\cite{WSLDAToolkit}.

When measuring the monopole mode frequency $\omega$, we perturb the ground state by multiplying it by a factor $\exp(-i\epsilon x^2/a^2)$, where $\epsilon$ is of the order of $10^{-6}$ in our simulations. Afterwards, we fit the droplet width $\sqrt{\langle x^2\rangle - \langle x\rangle^2}(t)$ to a function $f(t)=A+B\cos(\omega t + C)$, where $A$, $B$, $C$, and $\omega$ are fitted constants.

In order to estimate the uncertainty due to the quality of the energy density functional, we repeat the simulations with alternative spline representations of $\mathcal{E}_\mathrm{mLLGP}$. Namely we reduce the number of points we use to extrapolate the data with $f_L(n)$ and redo the whole procedure with a slightly different spline.

\section{{}\\Dark solitons in the mLLGP equation\label{app:solitons}}

To find the dark solitonic solutions of the mLLGP equation, we go to the thermodynamic limit, i.e. $L\to\infty, N\to\infty$ and $N/L=const$. We plug the following Ansatz for a wave travelling through the system at a constant velocity $v_s$, i.e. $\psi(x,t)=\tilde\psi(\zeta)$, where $\zeta=x-v_st$ is a comoving coordinate, to Eq.~\eqref{eq:mLLGPE} and obtain 
\begin{equation}
    \mu_s\tilde\psi-imv_s\tilde\psi'=-\frac{\hbar^2}{2m}\tilde\psi''+\mu_\mathrm{mLLGP}\left[|\tilde\psi|^2;g,\gud\right]\tilde\psi.
\end{equation}

If we assume that far away from the soliton, the density and phase are constant $\lim_{\zeta\to\infty}|\tilde\psi(\zeta)|^2=n_\infty$ and $\lim_{\zeta\to\infty}\arg\tilde\psi(\zeta)=\varphi_\infty$, we can find the value of the chemical potential $\mu_s=\mu_\mathrm{mLLGP}\left[n_\infty;g,\gud\right]$. Then, we rewrite the equation above in a discretized form, assuming that we start from two points far away from the soliton $\tilde\psi_0=(1-\epsilon_1)\sqrt{n_\infty}$ and $\tilde\psi_1=(1-\epsilon_2)\sqrt{n_\infty}$ with $\epsilon_{1,2}\ll1$ (typically $\sim 10^{-4}$) and $\epsilon_1>\epsilon_2$.

We have also checked
that solitonic solutions are dynamically stable.

\bibliography{apssamp}

\begin{thebibliography}{55}%
\makeatletter
\providecommand \@ifxundefined [1]{%
 \@ifx{#1\undefined}
}%
\providecommand \@ifnum [1]{%
 \ifnum #1\expandafter \@firstoftwo
 \else \expandafter \@secondoftwo
 \fi
}%
\providecommand \@ifx [1]{%
 \ifx #1\expandafter \@firstoftwo
 \else \expandafter \@secondoftwo
 \fi
}%
\providecommand \natexlab [1]{#1}%
\providecommand \enquote  [1]{``#1''}%
\providecommand \bibnamefont  [1]{#1}%
\providecommand \bibfnamefont [1]{#1}%
\providecommand \citenamefont [1]{#1}%
\providecommand \href@noop [0]{\@secondoftwo}%
\providecommand \href [0]{\begingroup \@sanitize@url \@href}%
\providecommand \@href[1]{\@@startlink{#1}\@@href}%
\providecommand \@@href[1]{\endgroup#1\@@endlink}%
\providecommand \@sanitize@url [0]{\catcode `\\12\catcode `\$12\catcode
  `\&12\catcode `\#12\catcode `\^12\catcode `\_12\catcode `\%12\relax}%
\providecommand \@@startlink[1]{}%
\providecommand \@@endlink[0]{}%
\providecommand \url  [0]{\begingroup\@sanitize@url \@url }%
\providecommand \@url [1]{\endgroup\@href {#1}{\urlprefix }}%
\providecommand \urlprefix  [0]{URL }%
\providecommand \Eprint [0]{\href }%
\providecommand \doibase [0]{https://doi.org/}%
\providecommand \selectlanguage [0]{\@gobble}%
\providecommand \bibinfo  [0]{\@secondoftwo}%
\providecommand \bibfield  [0]{\@secondoftwo}%
\providecommand \translation [1]{[#1]}%
\providecommand \BibitemOpen [0]{}%
\providecommand \bibitemStop [0]{}%
\providecommand \bibitemNoStop [0]{.\EOS\space}%
\providecommand \EOS [0]{\spacefactor3000\relax}%
\providecommand \BibitemShut  [1]{\csname bibitem#1\endcsname}%
\let\auto@bib@innerbib\@empty
%</preamble>
\bibitem [{\citenamefont {Bulgac}(2002)}]{Bulgac_2002}%
  \BibitemOpen
  \bibfield  {author} {\bibinfo {author} {\bibfnamefont {A.}~\bibnamefont
  {Bulgac}},\ }\bibfield  {title} {\bibinfo {title} {Dilute quantum droplets},\
  }\href {https://doi.org/10.1103/PhysRevLett.89.050402} {\bibfield  {journal}
  {\bibinfo  {journal} {Phys. Rev. Lett.}\ }\textbf {\bibinfo {volume} {89}},\
  \bibinfo {pages} {050402} (\bibinfo {year} {2002})}\BibitemShut {NoStop}%
\bibitem [{\citenamefont {Petrov}(2015)}]{Petrov_2015}%
  \BibitemOpen
  \bibfield  {author} {\bibinfo {author} {\bibfnamefont {D.~S.}\ \bibnamefont
  {Petrov}},\ }\bibfield  {title} {\bibinfo {title} {Quantum mechanical
  stabilization of a collapsing{ Bose-Bose} mixture},\ }\href
  {https://doi.org/10.1103/PhysRevLett.115.155302} {\bibfield  {journal}
  {\bibinfo  {journal} {Phys. Rev. Lett.}\ }\textbf {\bibinfo {volume} {115}},\
  \bibinfo {pages} {155302} (\bibinfo {year} {2015})}\BibitemShut {NoStop}%
\bibitem [{\citenamefont {Lee}\ \emph {et~al.}(1957)\citenamefont {Lee},
  \citenamefont {Huang},\ and\ \citenamefont {Yang}}]{Lee_1957a}%
  \BibitemOpen
  \bibfield  {author} {\bibinfo {author} {\bibfnamefont {T.~D.}\ \bibnamefont
  {Lee}}, \bibinfo {author} {\bibfnamefont {K.}~\bibnamefont {Huang}},\ and\
  \bibinfo {author} {\bibfnamefont {C.~N.}\ \bibnamefont {Yang}},\ }\bibfield
  {title} {\bibinfo {title} {Eigenvalues and eigenfunctions of a {Bose} system
  of hard spheres and its low-temperature properties},\ }\href
  {https://doi.org/10.1103/PhysRev.106.1135} {\bibfield  {journal} {\bibinfo
  {journal} {Phys. Rev.}\ }\textbf {\bibinfo {volume} {106}},\ \bibinfo {pages}
  {1135} (\bibinfo {year} {1957})}\BibitemShut {NoStop}%
\bibitem [{\citenamefont {Lee}\ and\ \citenamefont {Yang}(1957)}]{Lee_1957b}%
  \BibitemOpen
  \bibfield  {author} {\bibinfo {author} {\bibfnamefont {T.~D.}\ \bibnamefont
  {Lee}}\ and\ \bibinfo {author} {\bibfnamefont {C.~N.}\ \bibnamefont {Yang}},\
  }\bibfield  {title} {\bibinfo {title} {Many-body problem in quantum mechanics
  and quantum statistical mechanics},\ }\href
  {https://doi.org/10.1103/PhysRev.105.1119} {\bibfield  {journal} {\bibinfo
  {journal} {Phys. Rev.}\ }\textbf {\bibinfo {volume} {105}},\ \bibinfo {pages}
  {1119} (\bibinfo {year} {1957})}\BibitemShut {NoStop}%
\bibitem [{\citenamefont {Astrakharchik}\ and\ \citenamefont
  {Malomed}(2018)}]{Astrakharchik_2018}%
  \BibitemOpen
  \bibfield  {author} {\bibinfo {author} {\bibfnamefont {G.~E.}\ \bibnamefont
  {Astrakharchik}}\ and\ \bibinfo {author} {\bibfnamefont {B.~A.}\ \bibnamefont
  {Malomed}},\ }\bibfield  {title} {\bibinfo {title} {Dynamics of
  one-dimensional quantum droplets},\ }\href
  {https://doi.org/10.1103/PhysRevA.98.013631} {\bibfield  {journal} {\bibinfo
  {journal} {Phys. Rev. A}\ }\textbf {\bibinfo {volume} {98}},\ \bibinfo
  {pages} {013631} (\bibinfo {year} {2018})}\BibitemShut {NoStop}%
\bibitem [{\citenamefont {Petrov}\ and\ \citenamefont
  {Astrakharchik}(2016)}]{Petrov_2016}%
  \BibitemOpen
  \bibfield  {author} {\bibinfo {author} {\bibfnamefont {D.~S.}\ \bibnamefont
  {Petrov}}\ and\ \bibinfo {author} {\bibfnamefont {G.~E.}\ \bibnamefont
  {Astrakharchik}},\ }\bibfield  {title} {\bibinfo {title} {Ultradilute
  low-dimensional liquids},\ }\href
  {https://doi.org/10.1103/PhysRevLett.117.100401} {\bibfield  {journal}
  {\bibinfo  {journal} {Phys. Rev. Lett.}\ }\textbf {\bibinfo {volume} {117}},\
  \bibinfo {pages} {100401} (\bibinfo {year} {2016})}\BibitemShut {NoStop}%
\bibitem [{\citenamefont {Tylutki}\ \emph {et~al.}(2020)\citenamefont
  {Tylutki}, \citenamefont {Astrakharchik}, \citenamefont {Malomed},\ and\
  \citenamefont {Petrov}}]{Tylutki_2020}%
  \BibitemOpen
  \bibfield  {author} {\bibinfo {author} {\bibfnamefont {M.}~\bibnamefont
  {Tylutki}}, \bibinfo {author} {\bibfnamefont {G.~E.}\ \bibnamefont
  {Astrakharchik}}, \bibinfo {author} {\bibfnamefont {B.~A.}\ \bibnamefont
  {Malomed}},\ and\ \bibinfo {author} {\bibfnamefont {D.~S.}\ \bibnamefont
  {Petrov}},\ }\bibfield  {title} {\bibinfo {title} {Collective excitations of
  a one-dimensional quantum droplet},\ }\href
  {https://doi.org/10.1103/PhysRevA.101.051601} {\bibfield  {journal} {\bibinfo
   {journal} {Phys. Rev. A}\ }\textbf {\bibinfo {volume} {101}},\ \bibinfo
  {pages} {051601} (\bibinfo {year} {2020})}\BibitemShut {NoStop}%
\bibitem [{\citenamefont {Flynn}\ \emph {et~al.}(2022)\citenamefont {Flynn},
  \citenamefont {Parisi}, \citenamefont {Billam},\ and\ \citenamefont
  {Parker}}]{Flynn_2022}%
  \BibitemOpen
  \bibfield  {author} {\bibinfo {author} {\bibfnamefont {T.~A.}\ \bibnamefont
  {Flynn}}, \bibinfo {author} {\bibfnamefont {L.}~\bibnamefont {Parisi}},
  \bibinfo {author} {\bibfnamefont {T.~P.}\ \bibnamefont {Billam}},\ and\
  \bibinfo {author} {\bibfnamefont {N.~G.}\ \bibnamefont {Parker}},\
  }\href@noop {} {\bibinfo {title} {Quantum droplets in imbalanced atomic
  mixtures}} (\bibinfo {year} {2022}),\ \Eprint
  {https://arxiv.org/abs/2209.04318} {arXiv:2209.04318 [cond-mat.quant-gas]}
  \BibitemShut {NoStop}%
\bibitem [{\citenamefont {Cabrera}\ \emph {et~al.}(2018)\citenamefont
  {Cabrera}, \citenamefont {Tanzi}, \citenamefont {Sanz}, \citenamefont
  {Naylor}, \citenamefont {Thomas}, \citenamefont {Cheiney},\ and\
  \citenamefont {Tarruell}}]{Cabrera_2018}%
  \BibitemOpen
  \bibfield  {author} {\bibinfo {author} {\bibfnamefont {C.~R.}\ \bibnamefont
  {Cabrera}}, \bibinfo {author} {\bibfnamefont {L.}~\bibnamefont {Tanzi}},
  \bibinfo {author} {\bibfnamefont {J.}~\bibnamefont {Sanz}}, \bibinfo {author}
  {\bibfnamefont {B.}~\bibnamefont {Naylor}}, \bibinfo {author} {\bibfnamefont
  {P.}~\bibnamefont {Thomas}}, \bibinfo {author} {\bibfnamefont
  {P.}~\bibnamefont {Cheiney}},\ and\ \bibinfo {author} {\bibfnamefont
  {L.}~\bibnamefont {Tarruell}},\ }\bibfield  {title} {\bibinfo {title}
  {Quantum liquid droplets in a mixture of {Bose-Einstein} condensates},\
  }\href {https://doi.org/10.1126/science.aao5686} {\bibfield  {journal}
  {\bibinfo  {journal} {Science}\ }\textbf {\bibinfo {volume} {359}},\ \bibinfo
  {pages} {301} (\bibinfo {year} {2018})}\BibitemShut {NoStop}%
\bibitem [{\citenamefont {Cheiney}\ \emph {et~al.}(2018)\citenamefont
  {Cheiney}, \citenamefont {Cabrera}, \citenamefont {Sanz}, \citenamefont
  {Naylor}, \citenamefont {Tanzi},\ and\ \citenamefont
  {Tarruell}}]{Cheiney_2018}%
  \BibitemOpen
  \bibfield  {author} {\bibinfo {author} {\bibfnamefont {P.}~\bibnamefont
  {Cheiney}}, \bibinfo {author} {\bibfnamefont {C.~R.}\ \bibnamefont
  {Cabrera}}, \bibinfo {author} {\bibfnamefont {J.}~\bibnamefont {Sanz}},
  \bibinfo {author} {\bibfnamefont {B.}~\bibnamefont {Naylor}}, \bibinfo
  {author} {\bibfnamefont {L.}~\bibnamefont {Tanzi}},\ and\ \bibinfo {author}
  {\bibfnamefont {L.}~\bibnamefont {Tarruell}},\ }\bibfield  {title} {\bibinfo
  {title} {Bright soliton to quantum droplet transition in a mixture of
  {Bose-Einstein} condensates},\ }\href
  {https://doi.org/10.1103/PhysRevLett.120.135301} {\bibfield  {journal}
  {\bibinfo  {journal} {Phys. Rev. Lett.}\ }\textbf {\bibinfo {volume} {120}},\
  \bibinfo {pages} {135301} (\bibinfo {year} {2018})}\BibitemShut {NoStop}%
\bibitem [{\citenamefont {Semeghini}\ \emph {et~al.}(2018)\citenamefont
  {Semeghini}, \citenamefont {Ferioli}, \citenamefont {Masi}, \citenamefont
  {Mazzinghi}, \citenamefont {Wolswijk}, \citenamefont {Minardi}, \citenamefont
  {Modugno}, \citenamefont {Modugno}, \citenamefont {Inguscio},\ and\
  \citenamefont {Fattori}}]{Semeghini_2018}%
  \BibitemOpen
  \bibfield  {author} {\bibinfo {author} {\bibfnamefont {G.}~\bibnamefont
  {Semeghini}}, \bibinfo {author} {\bibfnamefont {G.}~\bibnamefont {Ferioli}},
  \bibinfo {author} {\bibfnamefont {L.}~\bibnamefont {Masi}}, \bibinfo {author}
  {\bibfnamefont {C.}~\bibnamefont {Mazzinghi}}, \bibinfo {author}
  {\bibfnamefont {L.}~\bibnamefont {Wolswijk}}, \bibinfo {author}
  {\bibfnamefont {F.}~\bibnamefont {Minardi}}, \bibinfo {author} {\bibfnamefont
  {M.}~\bibnamefont {Modugno}}, \bibinfo {author} {\bibfnamefont
  {G.}~\bibnamefont {Modugno}}, \bibinfo {author} {\bibfnamefont
  {M.}~\bibnamefont {Inguscio}},\ and\ \bibinfo {author} {\bibfnamefont
  {M.}~\bibnamefont {Fattori}},\ }\bibfield  {title} {\bibinfo {title}
  {Self-bound quantum droplets of atomic mixtures in free space},\ }\href
  {https://doi.org/10.1103/PhysRevLett.120.235301} {\bibfield  {journal}
  {\bibinfo  {journal} {Phys. Rev. Lett.}\ }\textbf {\bibinfo {volume} {120}},\
  \bibinfo {pages} {235301} (\bibinfo {year} {2018})}\BibitemShut {NoStop}%
\bibitem [{\citenamefont {Ferioli}\ \emph {et~al.}(2019)\citenamefont
  {Ferioli}, \citenamefont {Semeghini}, \citenamefont {Masi}, \citenamefont
  {Giusti}, \citenamefont {Modugno}, \citenamefont {Inguscio}, \citenamefont
  {Gallem\'{\i}}, \citenamefont {Recati},\ and\ \citenamefont
  {Fattori}}]{Ferioli_2019}%
  \BibitemOpen
  \bibfield  {author} {\bibinfo {author} {\bibfnamefont {G.}~\bibnamefont
  {Ferioli}}, \bibinfo {author} {\bibfnamefont {G.}~\bibnamefont {Semeghini}},
  \bibinfo {author} {\bibfnamefont {L.}~\bibnamefont {Masi}}, \bibinfo {author}
  {\bibfnamefont {G.}~\bibnamefont {Giusti}}, \bibinfo {author} {\bibfnamefont
  {G.}~\bibnamefont {Modugno}}, \bibinfo {author} {\bibfnamefont
  {M.}~\bibnamefont {Inguscio}}, \bibinfo {author} {\bibfnamefont
  {A.}~\bibnamefont {Gallem\'{\i}}}, \bibinfo {author} {\bibfnamefont
  {A.}~\bibnamefont {Recati}},\ and\ \bibinfo {author} {\bibfnamefont
  {M.}~\bibnamefont {Fattori}},\ }\bibfield  {title} {\bibinfo {title}
  {Collisions of self-bound quantum droplets},\ }\href
  {https://doi.org/10.1103/PhysRevLett.122.090401} {\bibfield  {journal}
  {\bibinfo  {journal} {Phys. Rev. Lett.}\ }\textbf {\bibinfo {volume} {122}},\
  \bibinfo {pages} {090401} (\bibinfo {year} {2019})}\BibitemShut {NoStop}%
\bibitem [{\citenamefont {D'Errico}\ \emph {et~al.}(2019)\citenamefont
  {D'Errico}, \citenamefont {Burchianti}, \citenamefont {Prevedelli},
  \citenamefont {Salasnich}, \citenamefont {Ancilotto}, \citenamefont
  {Modugno}, \citenamefont {Minardi},\ and\ \citenamefont
  {Fort}}]{DErrico_2019}%
  \BibitemOpen
  \bibfield  {author} {\bibinfo {author} {\bibfnamefont {C.}~\bibnamefont
  {D'Errico}}, \bibinfo {author} {\bibfnamefont {A.}~\bibnamefont
  {Burchianti}}, \bibinfo {author} {\bibfnamefont {M.}~\bibnamefont
  {Prevedelli}}, \bibinfo {author} {\bibfnamefont {L.}~\bibnamefont
  {Salasnich}}, \bibinfo {author} {\bibfnamefont {F.}~\bibnamefont
  {Ancilotto}}, \bibinfo {author} {\bibfnamefont {M.}~\bibnamefont {Modugno}},
  \bibinfo {author} {\bibfnamefont {F.}~\bibnamefont {Minardi}},\ and\ \bibinfo
  {author} {\bibfnamefont {C.}~\bibnamefont {Fort}},\ }\bibfield  {title}
  {\bibinfo {title} {Observation of quantum droplets in a heteronuclear bosonic
  mixture},\ }\href {https://doi.org/10.1103/PhysRevResearch.1.033155}
  {\bibfield  {journal} {\bibinfo  {journal} {Phys. Rev. Research}\ }\textbf
  {\bibinfo {volume} {1}},\ \bibinfo {pages} {033155} (\bibinfo {year}
  {2019})}\BibitemShut {NoStop}%
\bibitem [{\citenamefont {Bisset}\ \emph {et~al.}(2021)\citenamefont {Bisset},
  \citenamefont {Ardila Pe\~na},\ and\ \citenamefont {Santos}}]{Bisset_2021}%
  \BibitemOpen
  \bibfield  {author} {\bibinfo {author} {\bibfnamefont {R.~N.}\ \bibnamefont
  {Bisset}}, \bibinfo {author} {\bibfnamefont {L.~A.}\ \bibnamefont {Ardila
  Pe\~na}},\ and\ \bibinfo {author} {\bibfnamefont {L.}~\bibnamefont
  {Santos}},\ }\bibfield  {title} {\bibinfo {title} {Quantum droplets of
  dipolar mixtures},\ }\href {https://doi.org/10.1103/PhysRevLett.126.025301}
  {\bibfield  {journal} {\bibinfo  {journal} {Phys. Rev. Lett.}\ }\textbf
  {\bibinfo {volume} {126}},\ \bibinfo {pages} {025301} (\bibinfo {year}
  {2021})}\BibitemShut {NoStop}%
\bibitem [{\citenamefont {Trautmann}\ \emph {et~al.}(2018)\citenamefont
  {Trautmann}, \citenamefont {Ilzh\"ofer}, \citenamefont {Durastante},
  \citenamefont {Politi}, \citenamefont {Sohmen}, \citenamefont {Mark},\ and\
  \citenamefont {Ferlaino}}]{Trautmann_2018}%
  \BibitemOpen
  \bibfield  {author} {\bibinfo {author} {\bibfnamefont {A.}~\bibnamefont
  {Trautmann}}, \bibinfo {author} {\bibfnamefont {P.}~\bibnamefont
  {Ilzh\"ofer}}, \bibinfo {author} {\bibfnamefont {G.}~\bibnamefont
  {Durastante}}, \bibinfo {author} {\bibfnamefont {C.}~\bibnamefont {Politi}},
  \bibinfo {author} {\bibfnamefont {M.}~\bibnamefont {Sohmen}}, \bibinfo
  {author} {\bibfnamefont {M.~J.}\ \bibnamefont {Mark}},\ and\ \bibinfo
  {author} {\bibfnamefont {F.}~\bibnamefont {Ferlaino}},\ }\bibfield  {title}
  {\bibinfo {title} {Dipolar quantum mixtures of erbium and dysprosium atoms},\
  }\href {https://doi.org/10.1103/PhysRevLett.121.213601} {\bibfield  {journal}
  {\bibinfo  {journal} {Phys. Rev. Lett.}\ }\textbf {\bibinfo {volume} {121}},\
  \bibinfo {pages} {213601} (\bibinfo {year} {2018})}\BibitemShut {NoStop}%
\bibitem [{\citenamefont {Durastante}\ \emph {et~al.}(2020)\citenamefont
  {Durastante}, \citenamefont {Politi}, \citenamefont {Sohmen}, \citenamefont
  {Ilzh\"ofer}, \citenamefont {Mark}, \citenamefont {Norcia},\ and\
  \citenamefont {Ferlaino}}]{Durastante_2020}%
  \BibitemOpen
  \bibfield  {author} {\bibinfo {author} {\bibfnamefont {G.}~\bibnamefont
  {Durastante}}, \bibinfo {author} {\bibfnamefont {C.}~\bibnamefont {Politi}},
  \bibinfo {author} {\bibfnamefont {M.}~\bibnamefont {Sohmen}}, \bibinfo
  {author} {\bibfnamefont {P.}~\bibnamefont {Ilzh\"ofer}}, \bibinfo {author}
  {\bibfnamefont {M.~J.}\ \bibnamefont {Mark}}, \bibinfo {author}
  {\bibfnamefont {M.~A.}\ \bibnamefont {Norcia}},\ and\ \bibinfo {author}
  {\bibfnamefont {F.}~\bibnamefont {Ferlaino}},\ }\bibfield  {title} {\bibinfo
  {title} {Feshbach resonances in an erbium-dysprosium dipolar mixture},\
  }\href {https://doi.org/10.1103/PhysRevA.102.033330} {\bibfield  {journal}
  {\bibinfo  {journal} {Phys. Rev. A}\ }\textbf {\bibinfo {volume} {102}},\
  \bibinfo {pages} {033330} (\bibinfo {year} {2020})}\BibitemShut {NoStop}%
\bibitem [{\citenamefont {Parisi}\ \emph {et~al.}(2019)\citenamefont {Parisi},
  \citenamefont {Astrakharchik},\ and\ \citenamefont {Giorgini}}]{Parisi_2019}%
  \BibitemOpen
  \bibfield  {author} {\bibinfo {author} {\bibfnamefont {L.}~\bibnamefont
  {Parisi}}, \bibinfo {author} {\bibfnamefont {G.~E.}\ \bibnamefont
  {Astrakharchik}},\ and\ \bibinfo {author} {\bibfnamefont {S.}~\bibnamefont
  {Giorgini}},\ }\bibfield  {title} {\bibinfo {title} {Liquid state of
  one-dimensional {Bose} mixtures: A quantum {Monte Carlo} study},\ }\href
  {https://doi.org/10.1103/PhysRevLett.122.105302} {\bibfield  {journal}
  {\bibinfo  {journal} {Phys. Rev. Lett.}\ }\textbf {\bibinfo {volume} {122}},\
  \bibinfo {pages} {105302} (\bibinfo {year} {2019})}\BibitemShut {NoStop}%
\bibitem [{\citenamefont {Parisi}\ and\ \citenamefont
  {Giorgini}(2020)}]{Parisi_2020}%
  \BibitemOpen
  \bibfield  {author} {\bibinfo {author} {\bibfnamefont {L.}~\bibnamefont
  {Parisi}}\ and\ \bibinfo {author} {\bibfnamefont {S.}~\bibnamefont
  {Giorgini}},\ }\bibfield  {title} {\bibinfo {title} {Quantum droplets in
  one-dimensional {Bose} mixtures: A quantum {Monte Carlo} study},\ }\href
  {https://doi.org/10.1103/PhysRevA.102.023318} {\bibfield  {journal} {\bibinfo
   {journal} {Phys. Rev. A}\ }\textbf {\bibinfo {volume} {102}},\ \bibinfo
  {pages} {023318} (\bibinfo {year} {2020})}\BibitemShut {NoStop}%
\bibitem [{\citenamefont {Mermin}\ and\ \citenamefont
  {Wagner}(1966)}]{Mermin_1966}%
  \BibitemOpen
  \bibfield  {author} {\bibinfo {author} {\bibfnamefont {N.~D.}\ \bibnamefont
  {Mermin}}\ and\ \bibinfo {author} {\bibfnamefont {H.}~\bibnamefont
  {Wagner}},\ }\bibfield  {title} {\bibinfo {title} {Absence of ferromagnetism
  or antiferromagnetism in one- or two-dimensional isotropic heisenberg
  models},\ }\href {https://doi.org/10.1103/PhysRevLett.17.1133} {\bibfield
  {journal} {\bibinfo  {journal} {Phys. Rev. Lett.}\ }\textbf {\bibinfo
  {volume} {17}},\ \bibinfo {pages} {1133} (\bibinfo {year}
  {1966})}\BibitemShut {NoStop}%
\bibitem [{\citenamefont {Hohenberg}(1967)}]{Hohenberg_1967}%
  \BibitemOpen
  \bibfield  {author} {\bibinfo {author} {\bibfnamefont {P.~C.}\ \bibnamefont
  {Hohenberg}},\ }\bibfield  {title} {\bibinfo {title} {Existence of long-range
  order in one and two dimensions},\ }\href
  {https://doi.org/10.1103/PhysRev.158.383} {\bibfield  {journal} {\bibinfo
  {journal} {Phys. Rev.}\ }\textbf {\bibinfo {volume} {158}},\ \bibinfo {pages}
  {383} (\bibinfo {year} {1967})}\BibitemShut {NoStop}%
\bibitem [{\citenamefont {Lieb}\ and\ \citenamefont
  {Liniger}(1963)}]{Lieb_1963}%
  \BibitemOpen
  \bibfield  {author} {\bibinfo {author} {\bibfnamefont {E.~H.}\ \bibnamefont
  {Lieb}}\ and\ \bibinfo {author} {\bibfnamefont {W.}~\bibnamefont {Liniger}},\
  }\bibfield  {title} {\bibinfo {title} {Exact analysis of an interacting
  {B}ose gas. {I. T}he general solution and the ground state},\ }\href
  {https://doi.org/10.1103/PhysRev.130.1605} {\bibfield  {journal} {\bibinfo
  {journal} {Phys. Rev.}\ }\textbf {\bibinfo {volume} {130}},\ \bibinfo {pages}
  {1605} (\bibinfo {year} {1963})}\BibitemShut {NoStop}%
\bibitem [{\citenamefont {Lieb}(1963)}]{Lieb_1963b}%
  \BibitemOpen
  \bibfield  {author} {\bibinfo {author} {\bibfnamefont {E.~H.}\ \bibnamefont
  {Lieb}},\ }\bibfield  {title} {\bibinfo {title} {Exact analysis of an
  interacting {Bose gas. II.} the excitation spectrum},\ }\href
  {https://doi.org/10.1103/PhysRev.130.1616} {\bibfield  {journal} {\bibinfo
  {journal} {Phys. Rev.}\ }\textbf {\bibinfo {volume} {130}},\ \bibinfo {pages}
  {1616} (\bibinfo {year} {1963})}\BibitemShut {NoStop}%
\bibitem [{\citenamefont {Dunjko}\ \emph {et~al.}(2001)\citenamefont {Dunjko},
  \citenamefont {Lorent},\ and\ \citenamefont {Olshanii}}]{Dunjko_2001}%
  \BibitemOpen
  \bibfield  {author} {\bibinfo {author} {\bibfnamefont {V.}~\bibnamefont
  {Dunjko}}, \bibinfo {author} {\bibfnamefont {V.}~\bibnamefont {Lorent}},\
  and\ \bibinfo {author} {\bibfnamefont {M.}~\bibnamefont {Olshanii}},\
  }\bibfield  {title} {\bibinfo {title} {Bosons in cigar-shaped traps:
  {Thomas-Fermi} regime, {Tonks-Girardeau} regime, and in between},\ }\href
  {https://doi.org/10.1103/PhysRevLett.86.5413} {\bibfield  {journal} {\bibinfo
   {journal} {Phys. Rev. Lett.}\ }\textbf {\bibinfo {volume} {86}},\ \bibinfo
  {pages} {5413} (\bibinfo {year} {2001})}\BibitemShut {NoStop}%
\bibitem [{\citenamefont {\"Ohberg}\ and\ \citenamefont
  {Santos}(2002)}]{Ohberg_2002}%
  \BibitemOpen
  \bibfield  {author} {\bibinfo {author} {\bibfnamefont {P.}~\bibnamefont
  {\"Ohberg}}\ and\ \bibinfo {author} {\bibfnamefont {L.}~\bibnamefont
  {Santos}},\ }\bibfield  {title} {\bibinfo {title} {Dynamical transition from
  a quasi-one-dimensional {Bose-Einstein} condensate to a {Tonks-Girardeau}
  gas},\ }\href {https://doi.org/10.1103/PhysRevLett.89.240402} {\bibfield
  {journal} {\bibinfo  {journal} {Phys. Rev. Lett.}\ }\textbf {\bibinfo
  {volume} {89}},\ \bibinfo {pages} {240402} (\bibinfo {year}
  {2002})}\BibitemShut {NoStop}%
\bibitem [{\citenamefont {Kim}\ and\ \citenamefont {Zubarev}(2003)}]{Kim_2003}%
  \BibitemOpen
  \bibfield  {author} {\bibinfo {author} {\bibfnamefont {Y.~E.}\ \bibnamefont
  {Kim}}\ and\ \bibinfo {author} {\bibfnamefont {A.~L.}\ \bibnamefont
  {Zubarev}},\ }\bibfield  {title} {\bibinfo {title} {Density-functional theory
  of bosons in a trap},\ }\href {https://doi.org/10.1103/PhysRevA.67.015602}
  {\bibfield  {journal} {\bibinfo  {journal} {Phys. Rev. A}\ }\textbf {\bibinfo
  {volume} {67}},\ \bibinfo {pages} {015602} (\bibinfo {year}
  {2003})}\BibitemShut {NoStop}%
\bibitem [{\citenamefont {Damski}(2004)}]{Damski_2004}%
  \BibitemOpen
  \bibfield  {author} {\bibinfo {author} {\bibfnamefont {B.}~\bibnamefont
  {Damski}},\ }\bibfield  {title} {\bibinfo {title} {Formation of shock waves
  in a{ Bose-Einstein} condensate},\ }\href
  {https://doi.org/10.1103/PhysRevA.69.043610} {\bibfield  {journal} {\bibinfo
  {journal} {Phys. Rev. A}\ }\textbf {\bibinfo {volume} {69}},\ \bibinfo
  {pages} {043610} (\bibinfo {year} {2004})}\BibitemShut {NoStop}%
\bibitem [{\citenamefont {Damski}(2006)}]{Damski_2006}%
  \BibitemOpen
  \bibfield  {author} {\bibinfo {author} {\bibfnamefont {B.}~\bibnamefont
  {Damski}},\ }\bibfield  {title} {\bibinfo {title} {Shock waves in a
  one-dimensional {Bose gas: From a Bose-Einstein condensate to a Tonks} gas},\
  }\href {https://doi.org/10.1103/PhysRevA.73.043601} {\bibfield  {journal}
  {\bibinfo  {journal} {Phys. Rev. A}\ }\textbf {\bibinfo {volume} {73}},\
  \bibinfo {pages} {043601} (\bibinfo {year} {2006})}\BibitemShut {NoStop}%
\bibitem [{\citenamefont {Peotta}\ and\ \citenamefont
  {Ventra}(2014)}]{Peotta_2014}%
  \BibitemOpen
  \bibfield  {author} {\bibinfo {author} {\bibfnamefont {S.}~\bibnamefont
  {Peotta}}\ and\ \bibinfo {author} {\bibfnamefont {M.~D.}\ \bibnamefont
  {Ventra}},\ }\bibfield  {title} {\bibinfo {title} {Quantum shock waves and
  population inversion in collisions of ultracold atomic clouds},\ }\href
  {https://doi.org/10.1103/PhysRevA.89.013621} {\bibfield  {journal} {\bibinfo
  {journal} {Phys. Rev. A}\ }\textbf {\bibinfo {volume} {89}},\ \bibinfo
  {pages} {013621} (\bibinfo {year} {2014})}\BibitemShut {NoStop}%
\bibitem [{\citenamefont {Choi}\ \emph {et~al.}(2015)\citenamefont {Choi},
  \citenamefont {Dunjko}, \citenamefont {Zhang},\ and\ \citenamefont
  {Olshanii}}]{Choi_2015}%
  \BibitemOpen
  \bibfield  {author} {\bibinfo {author} {\bibfnamefont {S.}~\bibnamefont
  {Choi}}, \bibinfo {author} {\bibfnamefont {V.}~\bibnamefont {Dunjko}},
  \bibinfo {author} {\bibfnamefont {Z.~D.}\ \bibnamefont {Zhang}},\ and\
  \bibinfo {author} {\bibfnamefont {M.}~\bibnamefont {Olshanii}},\ }\bibfield
  {title} {\bibinfo {title} {Monopole excitations of a harmonically trapped
  one-dimensional {Bose} gas from the ideal gas to the {Tonks-Girardeau}
  regime},\ }\href {https://doi.org/10.1103/PhysRevLett.115.115302} {\bibfield
  {journal} {\bibinfo  {journal} {Phys. Rev. Lett.}\ }\textbf {\bibinfo
  {volume} {115}},\ \bibinfo {pages} {115302} (\bibinfo {year}
  {2015})}\BibitemShut {NoStop}%
\bibitem [{\citenamefont {Kopyci\ifmmode~\acute{n}\else \'{n}\fi{}ski}\ \emph
  {et~al.}(2023)\citenamefont {Kopyci\ifmmode~\acute{n}\else \'{n}\fi{}ski},
  \citenamefont {\L{}ebek}, \citenamefont {G\'orecki},\ and\ \citenamefont
  {Paw\l{}owski}}]{Kopycinski_2022b}%
  \BibitemOpen
  \bibfield  {author} {\bibinfo {author} {\bibfnamefont {J.}~\bibnamefont
  {Kopyci\ifmmode~\acute{n}\else \'{n}\fi{}ski}}, \bibinfo {author}
  {\bibfnamefont {M.}~\bibnamefont {\L{}ebek}}, \bibinfo {author}
  {\bibfnamefont {W.}~\bibnamefont {G\'orecki}},\ and\ \bibinfo {author}
  {\bibfnamefont {K.}~\bibnamefont {Paw\l{}owski}},\ }\bibfield  {title}
  {\bibinfo {title} {Ultrawide dark solitons and droplet-soliton coexistence in
  a dipolar bose gas with strong contact interactions},\ }\href
  {https://doi.org/10.1103/PhysRevLett.130.043401} {\bibfield  {journal}
  {\bibinfo  {journal} {Phys. Rev. Lett.}\ }\textbf {\bibinfo {volume} {130}},\
  \bibinfo {pages} {043401} (\bibinfo {year} {2023})}\BibitemShut {NoStop}%
\bibitem [{\citenamefont {O\l{}dziejewski}\ \emph {et~al.}(2020)\citenamefont
  {O\l{}dziejewski}, \citenamefont {G\'orecki}, \citenamefont {Paw\l{}owski},\
  and\ \citenamefont {Rz{a}\ifmmode~\dot{z}\else
  \.{z}ewski}}]{Oldziejewski_2020}%
  \BibitemOpen
  \bibfield  {author} {\bibinfo {author} {\bibfnamefont {R.}~\bibnamefont
  {O\l{}dziejewski}}, \bibinfo {author} {\bibfnamefont {W.}~\bibnamefont
  {G\'orecki}}, \bibinfo {author} {\bibfnamefont {K.}~\bibnamefont
  {Paw\l{}owski}},\ and\ \bibinfo {author} {\bibfnamefont {K.}~\bibnamefont
  {Rz{a}\ifmmode~\dot{z}\else \.{z}ewski}},\ }\bibfield  {title} {\bibinfo
  {title} {Strongly correlated quantum droplets in quasi-{1D} dipolar {Bose}
  gas},\ }\href {https://doi.org/10.1103/PhysRevLett.124.090401} {\bibfield
  {journal} {\bibinfo  {journal} {Phys. Rev. Lett.}\ }\textbf {\bibinfo
  {volume} {124}},\ \bibinfo {pages} {090401} (\bibinfo {year}
  {2020})}\BibitemShut {NoStop}%
\bibitem [{\citenamefont {De~Palo}\ \emph {et~al.}(2021)\citenamefont
  {De~Palo}, \citenamefont {Orignac}, \citenamefont {Chiofalo},\ and\
  \citenamefont {Citro}}]{DePalo_2021}%
  \BibitemOpen
  \bibfield  {author} {\bibinfo {author} {\bibfnamefont {S.}~\bibnamefont
  {De~Palo}}, \bibinfo {author} {\bibfnamefont {E.}~\bibnamefont {Orignac}},
  \bibinfo {author} {\bibfnamefont {M.~L.}\ \bibnamefont {Chiofalo}},\ and\
  \bibinfo {author} {\bibfnamefont {R.}~\bibnamefont {Citro}},\ }\bibfield
  {title} {\bibinfo {title} {Polarization angle dependence of the breathing
  mode in confined one-dimensional dipolar bosons},\ }\href
  {https://doi.org/10.1103/PhysRevB.103.115109} {\bibfield  {journal} {\bibinfo
   {journal} {Phys. Rev. B}\ }\textbf {\bibinfo {volume} {103}},\ \bibinfo
  {pages} {115109} (\bibinfo {year} {2021})}\BibitemShut {NoStop}%
\bibitem [{\citenamefont {Kopyciński}\ \emph {et~al.}(2022)\citenamefont
  {Kopyciński}, \citenamefont {Łebek}, \citenamefont {Marciniak},
  \citenamefont {Ołdziejewski}, \citenamefont {Górecki},\ and\ \citenamefont
  {Pawłowski}}]{Kopycinski_2022a}%
  \BibitemOpen
  \bibfield  {author} {\bibinfo {author} {\bibfnamefont {J.}~\bibnamefont
  {Kopyciński}}, \bibinfo {author} {\bibfnamefont {M.}~\bibnamefont {Łebek}},
  \bibinfo {author} {\bibfnamefont {M.}~\bibnamefont {Marciniak}}, \bibinfo
  {author} {\bibfnamefont {R.}~\bibnamefont {Ołdziejewski}}, \bibinfo {author}
  {\bibfnamefont {W.}~\bibnamefont {Górecki}},\ and\ \bibinfo {author}
  {\bibfnamefont {K.}~\bibnamefont {Pawłowski}},\ }\bibfield  {title}
  {\bibinfo {title} {{Beyond Gross-Pitaevskii equation for {1D} gas:
  quasiparticles and solitons}},\ }\href
  {https://doi.org/10.21468/SciPostPhys.12.1.023} {\bibfield  {journal}
  {\bibinfo  {journal} {SciPost Phys.}\ }\textbf {\bibinfo {volume} {12}},\
  \bibinfo {pages} {023} (\bibinfo {year} {2022})}\BibitemShut {NoStop}%
\bibitem [{\citenamefont {De~Palo}\ \emph {et~al.}(2022)\citenamefont
  {De~Palo}, \citenamefont {Orignac},\ and\ \citenamefont
  {Citro}}]{DePalo_2022}%
  \BibitemOpen
  \bibfield  {author} {\bibinfo {author} {\bibfnamefont {S.}~\bibnamefont
  {De~Palo}}, \bibinfo {author} {\bibfnamefont {E.}~\bibnamefont {Orignac}},\
  and\ \bibinfo {author} {\bibfnamefont {R.}~\bibnamefont {Citro}},\ }\bibfield
   {title} {\bibinfo {title} {Formation and fragmentation of quantum droplets
  in a quasi-one-dimensional dipolar {Bose} gas},\ }\href
  {https://doi.org/10.1103/PhysRevB.106.014503} {\bibfield  {journal} {\bibinfo
   {journal} {Phys. Rev. B}\ }\textbf {\bibinfo {volume} {106}},\ \bibinfo
  {pages} {014503} (\bibinfo {year} {2022})}\BibitemShut {NoStop}%
\bibitem [{\citenamefont {Łebek}\ \emph {et~al.}(2022)\citenamefont {Łebek},
  \citenamefont {Kopyciński}, \citenamefont {Górecki}, \citenamefont
  {Ołdziejewski},\ and\ \citenamefont {Pawłowski}}]{Lebek_2022}%
  \BibitemOpen
  \bibfield  {author} {\bibinfo {author} {\bibfnamefont {M.}~\bibnamefont
  {Łebek}}, \bibinfo {author} {\bibfnamefont {J.}~\bibnamefont {Kopyciński}},
  \bibinfo {author} {\bibfnamefont {W.}~\bibnamefont {Górecki}}, \bibinfo
  {author} {\bibfnamefont {R.}~\bibnamefont {Ołdziejewski}},\ and\ \bibinfo
  {author} {\bibfnamefont {K.}~\bibnamefont {Pawłowski}},\ }\href@noop {}
  {\bibinfo {title} {Elementary excitations of dipolar {Tonks-Girardeau}
  droplets}} (\bibinfo {year} {2022}),\ \Eprint
  {https://arxiv.org/abs/2209.01887} {arXiv:2209.01887 [cond-mat.quant-gas]}
  \BibitemShut {NoStop}%
\bibitem [{\citenamefont {Cikojevi\'{c}}\ \emph {et~al.}(2020)\citenamefont
  {Cikojevi\'{c}}, \citenamefont {Marki\'{c}}, \citenamefont {Pi},
  \citenamefont {Barranco},\ and\ \citenamefont {Boronat}}]{Cikojevic_2020}%
  \BibitemOpen
  \bibfield  {author} {\bibinfo {author} {\bibfnamefont {V.}~\bibnamefont
  {Cikojevi\'{c}}}, \bibinfo {author} {\bibfnamefont {L.~V.}\ \bibnamefont
  {Marki\'{c}}}, \bibinfo {author} {\bibfnamefont {M.}~\bibnamefont {Pi}},
  \bibinfo {author} {\bibfnamefont {M.}~\bibnamefont {Barranco}},\ and\
  \bibinfo {author} {\bibfnamefont {J.}~\bibnamefont {Boronat}},\ }\bibfield
  {title} {\bibinfo {title} {Towards a quantum {Monte Carlo}--based density
  functional including finite-range effects: Excitation modes of a
  $^{39}\mathrm{K}$ quantum droplet},\ }\href
  {https://doi.org/10.1103/PhysRevA.102.033335} {\bibfield  {journal} {\bibinfo
   {journal} {Phys. Rev. A}\ }\textbf {\bibinfo {volume} {102}},\ \bibinfo
  {pages} {033335} (\bibinfo {year} {2020})}\BibitemShut {NoStop}%
\bibitem [{\citenamefont {Hu}\ \emph {et~al.}(2020)\citenamefont {Hu},
  \citenamefont {Wang},\ and\ \citenamefont {Liu}}]{Hu_2020}%
  \BibitemOpen
  \bibfield  {author} {\bibinfo {author} {\bibfnamefont {H.}~\bibnamefont
  {Hu}}, \bibinfo {author} {\bibfnamefont {J.}~\bibnamefont {Wang}},\ and\
  \bibinfo {author} {\bibfnamefont {X.-J.}\ \bibnamefont {Liu}},\ }\bibfield
  {title} {\bibinfo {title} {Microscopic pairing theory of a binary bose
  mixture with interspecies attractions: {Bosonic BEC-BCS} crossover and
  ultradilute low-dimensional quantum droplets},\ }\href
  {https://doi.org/10.1103/PhysRevA.102.043301} {\bibfield  {journal} {\bibinfo
   {journal} {Phys. Rev. A}\ }\textbf {\bibinfo {volume} {102}},\ \bibinfo
  {pages} {043301} (\bibinfo {year} {2020})}\BibitemShut {NoStop}%
\bibitem [{\citenamefont {Ota}\ and\ \citenamefont
  {Astrakharchik}(2020)}]{Ota_2020}%
  \BibitemOpen
  \bibfield  {author} {\bibinfo {author} {\bibfnamefont {M.}~\bibnamefont
  {Ota}}\ and\ \bibinfo {author} {\bibfnamefont {G.~E.}\ \bibnamefont
  {Astrakharchik}},\ }\bibfield  {title} {\bibinfo {title} {{Beyond
  Lee-Huang-Yang description of self-bound Bose mixtures}},\ }\href
  {https://doi.org/10.21468/SciPostPhys.9.2.020} {\bibfield  {journal}
  {\bibinfo  {journal} {SciPost Phys.}\ }\textbf {\bibinfo {volume} {9}},\
  \bibinfo {pages} {020} (\bibinfo {year} {2020})}\BibitemShut {NoStop}%
\bibitem [{\citenamefont {Fiolhais}\ \emph {et~al.}(2003)\citenamefont
  {Fiolhais}, \citenamefont {Nogueira},\ and\ \citenamefont
  {Marques}}]{Fiolhais_2003}%
  \BibitemOpen
  \bibinfo {editor} {\bibfnamefont {C.}~\bibnamefont {Fiolhais}}, \bibinfo
  {editor} {\bibfnamefont {F.}~\bibnamefont {Nogueira}},\ and\ \bibinfo
  {editor} {\bibfnamefont {M.}~\bibnamefont {Marques}},\ eds.,\ \href@noop {}
  {\emph {\bibinfo {title} {A primer in density functional theory"}}},\
  \bibinfo {edition} {2003rd}\ ed.,\ Lecture Notes in Physics\ (\bibinfo
  {publisher} {Springer},\ \bibinfo {address} {Berlin, Germany},\ \bibinfo
  {year} {2003})\BibitemShut {NoStop}%
\bibitem [{\citenamefont {Bulgac}\ \emph {et~al.}(2012)\citenamefont {Bulgac},
  \citenamefont {Forbes},\ and\ \citenamefont {Magierski}}]{Bulgac_2012a}%
  \BibitemOpen
  \bibfield  {author} {\bibinfo {author} {\bibfnamefont {A.}~\bibnamefont
  {Bulgac}}, \bibinfo {author} {\bibfnamefont {M.~M.}\ \bibnamefont {Forbes}},\
  and\ \bibinfo {author} {\bibfnamefont {P.}~\bibnamefont {Magierski}},\
  }\bibinfo {title} {The unitary {F}ermi gas: From {Monte Carlo} to density
  functionals},\ in\ \href {https://doi.org/10.1007/978-3-642-21978-8_9} {\emph
  {\bibinfo {booktitle} {The BCS-BEC Crossover and the Unitary Fermi Gas}}},\
  \bibinfo {editor} {edited by\ \bibinfo {editor} {\bibfnamefont
  {W.}~\bibnamefont {Zwerger}}}\ (\bibinfo  {publisher} {Springer Berlin
  Heidelberg},\ \bibinfo {address} {Berlin, Heidelberg},\ \bibinfo {year}
  {2012})\ pp.\ \bibinfo {pages} {305--373}\BibitemShut {NoStop}%
\bibitem [{\citenamefont {Bulgac}\ and\ \citenamefont
  {Yoon}(2009)}]{Bulgac_2009}%
  \BibitemOpen
  \bibfield  {author} {\bibinfo {author} {\bibfnamefont {A.}~\bibnamefont
  {Bulgac}}\ and\ \bibinfo {author} {\bibfnamefont {S.}~\bibnamefont {Yoon}},\
  }\bibfield  {title} {\bibinfo {title} {Large amplitude dynamics of the
  pairing correlations in a unitary {Fermi} gas},\ }\href
  {https://doi.org/10.1103/PhysRevLett.102.085302} {\bibfield  {journal}
  {\bibinfo  {journal} {Phys. Rev. Lett.}\ }\textbf {\bibinfo {volume} {102}},\
  \bibinfo {pages} {085302} (\bibinfo {year} {2009})}\BibitemShut {NoStop}%
\bibitem [{\citenamefont {Wlaz\l{}owski}\ \emph {et~al.}(2018)\citenamefont
  {Wlaz\l{}owski}, \citenamefont {Sekizawa}, \citenamefont {Marchwiany},\ and\
  \citenamefont {Magierski}}]{Wlazlowski_2018}%
  \BibitemOpen
  \bibfield  {author} {\bibinfo {author} {\bibfnamefont {G.}~\bibnamefont
  {Wlaz\l{}owski}}, \bibinfo {author} {\bibfnamefont {K.}~\bibnamefont
  {Sekizawa}}, \bibinfo {author} {\bibfnamefont {M.}~\bibnamefont
  {Marchwiany}},\ and\ \bibinfo {author} {\bibfnamefont {P.}~\bibnamefont
  {Magierski}},\ }\bibfield  {title} {\bibinfo {title} {Suppressed solitonic
  cascade in spin-imbalanced superfluid {Fermi} gas},\ }\href
  {https://doi.org/10.1103/PhysRevLett.120.253002} {\bibfield  {journal}
  {\bibinfo  {journal} {Phys. Rev. Lett.}\ }\textbf {\bibinfo {volume} {120}},\
  \bibinfo {pages} {253002} (\bibinfo {year} {2018})}\BibitemShut {NoStop}%
\bibitem [{\citenamefont {Magierski}\ \emph {et~al.}(2019)\citenamefont
  {Magierski}, \citenamefont {T\"uzemen},\ and\ \citenamefont
  {Wlaz\l{}owski}}]{Magierski_2019}%
  \BibitemOpen
  \bibfield  {author} {\bibinfo {author} {\bibfnamefont {P.}~\bibnamefont
  {Magierski}}, \bibinfo {author} {\bibfnamefont {B.}~\bibnamefont
  {T\"uzemen}},\ and\ \bibinfo {author} {\bibfnamefont {G.}~\bibnamefont
  {Wlaz\l{}owski}},\ }\bibfield  {title} {\bibinfo {title} {Spin-polarized
  droplets in the unitary {Fermi} gas},\ }\href
  {https://doi.org/10.1103/PhysRevA.100.033613} {\bibfield  {journal} {\bibinfo
   {journal} {Phys. Rev. A}\ }\textbf {\bibinfo {volume} {100}},\ \bibinfo
  {pages} {033613} (\bibinfo {year} {2019})}\BibitemShut {NoStop}%
\bibitem [{\citenamefont {Tylutki}\ and\ \citenamefont
  {Wlaz\l{}owski}(2021)}]{Tylutki_2021}%
  \BibitemOpen
  \bibfield  {author} {\bibinfo {author} {\bibfnamefont {M.}~\bibnamefont
  {Tylutki}}\ and\ \bibinfo {author} {\bibfnamefont {G.}~\bibnamefont
  {Wlaz\l{}owski}},\ }\bibfield  {title} {\bibinfo {title} {Universal aspects
  of vortex reconnections across the {BCS-BEC} crossover},\ }\href
  {https://doi.org/10.1103/PhysRevA.103.L051302} {\bibfield  {journal}
  {\bibinfo  {journal} {Phys. Rev. A}\ }\textbf {\bibinfo {volume} {103}},\
  \bibinfo {pages} {L051302} (\bibinfo {year} {2021})}\BibitemShut {NoStop}%
\bibitem [{\citenamefont {Magierski}\ \emph {et~al.}(2022)\citenamefont
  {Magierski}, \citenamefont {Wlaz\l{}owski}, \citenamefont {Makowski},\ and\
  \citenamefont {Kobuszewski}}]{Magierski_2022}%
  \BibitemOpen
  \bibfield  {author} {\bibinfo {author} {\bibfnamefont {P.}~\bibnamefont
  {Magierski}}, \bibinfo {author} {\bibfnamefont {G.}~\bibnamefont
  {Wlaz\l{}owski}}, \bibinfo {author} {\bibfnamefont {A.}~\bibnamefont
  {Makowski}},\ and\ \bibinfo {author} {\bibfnamefont {K.}~\bibnamefont
  {Kobuszewski}},\ }\bibfield  {title} {\bibinfo {title} {Spin-polarized
  vortices with reversed circulation},\ }\href
  {https://doi.org/10.1103/PhysRevA.106.033322} {\bibfield  {journal} {\bibinfo
   {journal} {Phys. Rev. A}\ }\textbf {\bibinfo {volume} {106}},\ \bibinfo
  {pages} {033322} (\bibinfo {year} {2022})}\BibitemShut {NoStop}%
\bibitem [{\citenamefont {Zwierlein}\ \emph {et~al.}(2006)\citenamefont
  {Zwierlein}, \citenamefont {Schirotzek}, \citenamefont {Schunck},\ and\
  \citenamefont {Ketterle}}]{Zwierlein_2006}%
  \BibitemOpen
  \bibfield  {author} {\bibinfo {author} {\bibfnamefont {M.~W.}\ \bibnamefont
  {Zwierlein}}, \bibinfo {author} {\bibfnamefont {A.}~\bibnamefont
  {Schirotzek}}, \bibinfo {author} {\bibfnamefont {C.~H.}\ \bibnamefont
  {Schunck}},\ and\ \bibinfo {author} {\bibfnamefont {W.}~\bibnamefont
  {Ketterle}},\ }\bibfield  {title} {\bibinfo {title} {Fermionic superfluidity
  with imbalanced spin populations},\ }\href
  {https://doi.org/10.1126/science.1122318} {\bibfield  {journal} {\bibinfo
  {journal} {Science}\ }\textbf {\bibinfo {volume} {311}},\ \bibinfo {pages}
  {492} (\bibinfo {year} {2006})}\BibitemShut {NoStop}%
\bibitem [{\citenamefont {Ku}\ \emph {et~al.}(2016)\citenamefont {Ku},
  \citenamefont {Mukherjee}, \citenamefont {Yefsah},\ and\ \citenamefont
  {Zwierlein}}]{Ku_2016}%
  \BibitemOpen
  \bibfield  {author} {\bibinfo {author} {\bibfnamefont {M.~J.~H.}\
  \bibnamefont {Ku}}, \bibinfo {author} {\bibfnamefont {B.}~\bibnamefont
  {Mukherjee}}, \bibinfo {author} {\bibfnamefont {T.}~\bibnamefont {Yefsah}},\
  and\ \bibinfo {author} {\bibfnamefont {M.~W.}\ \bibnamefont {Zwierlein}},\
  }\bibfield  {title} {\bibinfo {title} {Cascade of solitonic excitations in a
  superfluid {Fermi} gas: From planar solitons to vortex rings and lines},\
  }\href {https://doi.org/10.1103/PhysRevLett.116.045304} {\bibfield  {journal}
  {\bibinfo  {journal} {Phys. Rev. Lett.}\ }\textbf {\bibinfo {volume} {116}},\
  \bibinfo {pages} {045304} (\bibinfo {year} {2016})}\BibitemShut {NoStop}%
\bibitem [{\citenamefont {Jackson}\ \emph {et~al.}(1998)\citenamefont
  {Jackson}, \citenamefont {Kavoulakis},\ and\ \citenamefont
  {Pethick}}]{Jackson_1998}%
  \BibitemOpen
  \bibfield  {author} {\bibinfo {author} {\bibfnamefont {A.~D.}\ \bibnamefont
  {Jackson}}, \bibinfo {author} {\bibfnamefont {G.~M.}\ \bibnamefont
  {Kavoulakis}},\ and\ \bibinfo {author} {\bibfnamefont {C.~J.}\ \bibnamefont
  {Pethick}},\ }\bibfield  {title} {\bibinfo {title} {Solitary waves in clouds
  of {Bose-Einstein} condensed atoms},\ }\href
  {https://doi.org/10.1103/PhysRevA.58.2417} {\bibfield  {journal} {\bibinfo
  {journal} {Phys. Rev. A}\ }\textbf {\bibinfo {volume} {58}},\ \bibinfo
  {pages} {2417} (\bibinfo {year} {1998})}\BibitemShut {NoStop}%
\bibitem [{\citenamefont {Morera}\ \emph {et~al.}(2018)\citenamefont {Morera},
  \citenamefont {Mateo}, \citenamefont {Polls},\ and\ \citenamefont
  {Juli\'a-D\'{\i}az}}]{Morera_2018}%
  \BibitemOpen
  \bibfield  {author} {\bibinfo {author} {\bibfnamefont {I.}~\bibnamefont
  {Morera}}, \bibinfo {author} {\bibfnamefont {A.~M.}\ \bibnamefont {Mateo}},
  \bibinfo {author} {\bibfnamefont {A.}~\bibnamefont {Polls}},\ and\ \bibinfo
  {author} {\bibfnamefont {B.}~\bibnamefont {Juli\'a-D\'{\i}az}},\ }\bibfield
  {title} {\bibinfo {title} {Dark-dark-soliton dynamics in two density-coupled
  {Bose-Einstein} condensates},\ }\href
  {https://doi.org/10.1103/PhysRevA.97.043621} {\bibfield  {journal} {\bibinfo
  {journal} {Phys. Rev. A}\ }\textbf {\bibinfo {volume} {97}},\ \bibinfo
  {pages} {043621} (\bibinfo {year} {2018})}\BibitemShut {NoStop}%
\bibitem [{\citenamefont {Kevrekidis}\ and\ \citenamefont
  {Frantzeskakis}(2016)}]{Keverekidis_2016}%
  \BibitemOpen
  \bibfield  {author} {\bibinfo {author} {\bibfnamefont {P.}~\bibnamefont
  {Kevrekidis}}\ and\ \bibinfo {author} {\bibfnamefont {D.}~\bibnamefont
  {Frantzeskakis}},\ }\bibfield  {title} {\bibinfo {title} {Solitons in coupled
  nonlinear {Schrödinger} models: {A} survey of recent developments},\ }\href
  {https://doi.org/https://doi.org/10.1016/j.revip.2016.07.002} {\bibfield
  {journal} {\bibinfo  {journal} {Reviews in Physics}\ }\textbf {\bibinfo
  {volume} {1}},\ \bibinfo {pages} {140} (\bibinfo {year} {2016})}\BibitemShut
  {NoStop}%
\bibitem [{\citenamefont {Edmonds}(2022)}]{Edmonds_2022}%
  \BibitemOpen
  \bibfield  {author} {\bibinfo {author} {\bibfnamefont {M.}~\bibnamefont
  {Edmonds}},\ }\href {https://doi.org/10.48550/ARXIV.2209.00790} {\bibinfo
  {title} {Dark quantum droplets in beyond-mean-field {Bose-Einstein}
  condensate mixtures}} (\bibinfo {year} {2022})\BibitemShut {NoStop}%
\bibitem [{Note1()}]{Note1}%
  \BibitemOpen
  \bibinfo {note} {Just as the original Lieb-Liniger Gross-Pitaevskii equation
  gives a quantitative agreement with the Lieb-Liniger model -- the homogeneous
  system energy, chemical potential and speed of sound are the same from
  construction.}\BibitemShut {Stop}%
\bibitem [{\citenamefont {De~Rosi}\ \emph {et~al.}(2021)\citenamefont
  {De~Rosi}, \citenamefont {Astrakharchik},\ and\ \citenamefont
  {Massignan}}]{DeRosi_2021}%
  \BibitemOpen
  \bibfield  {author} {\bibinfo {author} {\bibfnamefont {G.}~\bibnamefont
  {De~Rosi}}, \bibinfo {author} {\bibfnamefont {G.~E.}\ \bibnamefont
  {Astrakharchik}},\ and\ \bibinfo {author} {\bibfnamefont {P.}~\bibnamefont
  {Massignan}},\ }\bibfield  {title} {\bibinfo {title} {Thermal instability,
  evaporation, and thermodynamics of one-dimensional liquids in weakly
  interacting {Bose-Bose} mixtures},\ }\href
  {https://doi.org/10.1103/PhysRevA.103.043316} {\bibfield  {journal} {\bibinfo
   {journal} {Phys. Rev. A}\ }\textbf {\bibinfo {volume} {103}},\ \bibinfo
  {pages} {043316} (\bibinfo {year} {2021})}\BibitemShut {NoStop}%
\bibitem [{Note2()}]{Note2}%
  \BibitemOpen
  \bibinfo {note} {It is a benchmark of a correctly prepared fit.}\BibitemShut
  {Stop}%
\bibitem [{WSL()}]{WSLDAToolkit}%
  \BibitemOpen
  \href {https://wslda.fizyka.pw.edu.pl/} {\bibinfo {title} {{W-SLDA
  Toolkit}}},\ \bibinfo {howpublished}
  {\url{https://wslda.fizyka.pw.edu.pl/}}\BibitemShut {NoStop}%
\end{thebibliography}%

\end{document}